\documentclass[11pt,letterpaper]{article}
\usepackage{eurosym}
\usepackage{setspace}
\usepackage{color}
\usepackage{amsmath}
\usepackage{amsfonts}
\usepackage{subcaption}
\usepackage{comment}
\usepackage{booktabs}
\usepackage{verbatim}
\usepackage{amssymb}
\usepackage{graphicx}

\providecommand{\U}[1]{\protect\rule{.1in}{.1in}}
\begin{document}

\title{Crystals of gauged solitons, force free plasma and resurgence}
\author{Gonzalo Barriga$^{1,2}$, Fabrizio Canfora$^{1}$, Mat\'ias Torres$^{1,2}$, Aldo
Vera$^{3}$\\$^{1}$\textit{Centro de Estudios Cient\'{\i}ficos (CECS), Casilla 1469,
Valdivia, Chile,}\\$^{2}$\textit{Departamento de F\'{\i}sica, Universidad de Concepci\'{o}n,
Casilla 160-C, Concepci\'{o}n, Chile,}\\$^{3}$\textit{Instituto de Ciencias F\'isicas y Matem\'aticas, Universidad
Austral de Chile, Valdivia, Chile.}\\{\small gobarriga@udec.cl, canfora@cecs.cl, matiatorres@udec.cl,
aldo.vera@uach.cl}}
\maketitle
\begin{abstract}
We show that the (3+1)-dimensional gauged non-linear sigma model minimally
coupled to a $U(1)$ gauge field possesses analytic solutions representing
gauged solitons at finite Baryon density whose electromagnetic field is a
Force Free Plasma. These gauged solitons manifest a crystalline structure and
generate in a very natural way persistent currents able to support Force Free
Plasma electromagnetic fields. The trajectories of charged test particles
moving within these configurations can be characterized. Quite surprisingly,
despite the non-integrable nature of the theory, some of the perturbations of
these gauged solitons allow to identify a proper resurgent parameter. In
particular, the perturbations of the solitons profile are related to the
Lam\'{e} operator. On the other hand, the electromagnetic perturbations on the configurations 
satisfy a two-dimensional effective Schr\"{o}dinger equation,
where the soliton's background interacts with the electromagnetic
perturbations through an effective two-dimensional periodic potential. We
studied numerically the band energy spectrum for different values of the free
parameters of the theory and we found that bands-gaps are modulated by the
potential strength. Finally we compare our crystal solutions with those of the
(1+1)-dimesional Gross-Neveu model.

\end{abstract}

\newpage

\tableofcontents


\section{Introduction}


One of the most important types of plasma in plasma physics is the so-called
Force Free Plasma (FFP henceforth), and several reasons justify this
statement. First of all, FFP are extremely relevant in many astrophysical
situations. For instance, objects with extremely intense magnetic fields like
pulsars are typical sources of FFP. Secondly, FFP can be characterized in a
very elegant way with a non-linear set of PDEs for the electromagnetic field.
This is the so-called force free condition,
\begin{equation}
F_{\mu\nu}\nabla_{\rho}F^{\nu\rho}=0\ . \label{ffc}%
\end{equation}
The above equation is usually the starting point of the theoretical analysis
on FFP (see for instance \cite{WS2012}, \cite{FFP2} and references therein).
From one point of view, the above system of equations is very convenient since
many features of FFP can be analyzed without taking into account the
electromagnetic sources (which are absent in Eq. (\ref{ffc})). On the other
hand, this implies that often the very important physical question about which
are the actual concrete sources able to generate such FFP is somehow
neglected. In the present paper, as it will be explained in the following
sections, we identify a very natural and concrete source of FFP in the low
energy limit of Quantum Chromodynamics (QCD). FFP play also a key role in the
so-called \textit{Blandford-Znajek process }\cite{BZ}, mechanism that allows
to extract rotational energy from a spinning black hole. This happens when the
electric potential becomes larger than the threshold of electron-positron pair
creation, so that the black hole will be surrounded by a FFP in such a way
that the rotation energy can be radiated away. The pioneering theoretical
analysis in \cite{Carter:1979}, \cite{DT}, \cite{Uchida1}, \cite{Uchida2},
\cite{Uchida3}, \cite{Uchida4}, \cite{Uchida5}, \cite{Okamoto}, \cite{LGATN}
and \cite{GJ1} shed considerable light on this process.

In recent years relevant analytic examples of FFP have been constructed (see,
for instance, \cite{BGJ}, \cite{LRS}, \cite{BG}, \cite{YZ}, \cite{ZYL},
\cite{LR}, \cite{ZMP}, \cite{GJ2}, \cite{CO}, \cite{GLR} and references
therein), however, many of the available examples (which analyze in detail
solutions of the system in Eq. (\ref{ffc})) leave open the issue to identify
the actual sources of the force-free electromagnetic field. Due to the
interesting astrophysical applications of FFP it is of great importance to
find also analytic examples of FFP in which the sources of the electromagnetic
field can be traced back directly to observed particles of the standard model.
\textit{One of the goals of the present paper is to construct explicitly such
solutions}. In the present approach, the $SU(2)$ non-linear sigma model (NLSM
henceforth), which is the low limit of QCD describing Pions dynamic, provides
with explicit and topologically non-trivial sources of FFP. This is in
agreement with the fact that FFP must occur for objects made of Hadronic
matter (such as pulsars).

The NLSM is one of the most interesting effective field theories. Besides the
well-known relation with the low energy limit of QCD, such a model is also
very useful in statistical mechanics, in the theory of the quantum hall
effect, in the analysis of superfluids and so on (see \cite{manton},
\cite{BaMa}, \cite{2} and \cite{3}). It is worth to emphasize that the NLSM on
flat space-time does not possess static solutions with non-trivial topological
charge due to Derrick's scaling argument \cite{4}. A well-known method to
avoid this no-go theorem was found by Skyrme (even before the actual paper of
Derrick): the so-called Skyrme term \cite{skyrme} together with the NLSM
allows the existence of static solitonic solutions with finite energy and
topological charge, which represents the Baryonic charge, called Skyrmions
(see \cite{witten0}, \cite{bala0}, \cite{Bala1}, \cite{ANW} and references
therein). However, the original arguments in \cite{witten0} to prove that the
topological charge must be identified with the Baryonic charge do not require
explicitly the presence of the Skyrme term. Moreover, the Skyrme term is not
the only way to avoid the Derrick's argument. The Derrick's scaling argument
can be avoided by \textit{constructing time-dependent ansatz for the matter
fields with the characteristic that the corresponding energy-density is
stationary}. Hence, as we will show in the next sections, the minimal 
ingredients to build FFP together with the corresponding sources are already
present in the NLSM minimally coupled with Maxwell field.

The present strategy is based on the techniques developed in \cite{canfora2},
\cite{56}, \cite{56b}, \cite{58}, \cite{58b}, \cite{ACZ}, \cite{CanTalSk1},
\cite{canfora10}, \cite{gaugsk}, \cite{Canfora:2018clt},
\cite{Giacomini:2017xno}, \cite{ACLV}, \cite{lastEPJC}, \cite{Ayon2},
\cite{lastEPJCbis}, \cite{Canfora:2021rdw}, that have allowed to construct exact NLSM solutions with
non-trivial topological charge at finite density in \cite{Fab1},
\cite{crystal1}, \cite{crystal2}, \cite{crystal2.1}, \cite{CDP} and
\cite{CLV}. The gauged solitons constructed in \cite{crystal2} and 
\cite{crystal2.1} live at finite Baryon density and most of the energy and
the topological charge are contained within tube-shaped regions which are
regularly spaced\footnote{Exact crystalline solutions of ordered solitons with vanishing Baryonic charge have been constructed in \cite{Yamamoto1} and \cite{Yamamoto2}. Interestingly, the excitations around this chiral soliton lattice are related to the Lam\'e parameter.}. These structures are expected in the description of cold and
dense nuclear matter as a function of topological charge, commonly called
\textit{nuclear pasta}, and therefore are quite relevant in the phase diagram
of the low energy limit of QCD (see \cite{R1} and \cite{R2}). Although until
very recently the nuclear pasta phase was considered to be a very hard nut to
crack from the analytic viewpoint, there are very strong observational
evidences supporting it (see \cite{pasta1}, \cite{pasta2}, \cite{pasta3},
\cite{pasta4}\ and references therein). The comparison of the energy density
and Baryon density contour plots in \cite{crystal1}, \cite{crystal2} and
\cite{CLV} with the ones in, for instance, \cite{pasta3} is very encouraging.
Not only the analytic plots of energy density and Baryon density in
\cite{crystal1}, \cite{crystal2} and \cite{CLV} are very close to the
phenomenological ones in \cite{pasta1}, \cite{pasta2}, \cite{pasta3},
\cite{pasta4}, the present analytic framework also allows the explicit
computation of relevant quantities (such as the computation of the shear
modulus of the nuclear lasagna in \cite{CDP} which is close to the value of
the shear modulus of nuclear lasagna found in \cite{pasta5}). Here we will
show that the electromagnetic field generated by the solitons crystals found
in \cite{crystal1}, \cite{crystal2} (which describe quite well many features
of the \textit{nuclear spaghetti phase}) is of force free type. This
connection between FFP and the nuclear pasta phase is a quite surprising
result which (to the best of our knowledge) has not been discussed previously
in the literature. As it will be discussed in the next sections, this allows
among other things to discuss the trajectories of charged particles close to
these gauged solitons.

Another novel result of the present analysis is that some of the main physical
properties of these gauged solitons, which are natural sources of FFP,
manifest a ``\textit{resurgent character}". \textit{Resurgence} is one of the
few theoretical techniques which can help to make sense of the ubiquitous
factorially divergent series appearing in perturbation theory \cite{lecture1}.
Such strategy works as follows: First of all, one must use Borel summation in
the complex-$g$ plane ($g$ being the coupling constant of the theory) to
handle the given factorially divergent series so that the divergent series
becomes a finite expression. At a first glance, it seems that we have just
traded one problem for another (maybe worse) problem since the Borel sum
possesses ambiguities which manifest themselves along suitable lines in the
complex $g$-plane (for nice reviews see \cite{resurgence1} and
\cite{resurgence2}). However, when the models under investigation possess
suitable topological charges labelling non-trivial non-perturbative sectors,
the perturbative expansions in these topologically non-trivial sectors (which,
usually, are also factorially divergent) allow a remarkable rescue. In these
relevant theories, the ambiguities in the Borel summation of the perturbative
expansions in the topologically non-trivial sectors cancel the ambiguities of
the perturbative sector giving rise to a well-defined result \cite{[10]},
\cite{[11]}. Starting from \cite{[12]} in recent years there has been a great
revival of these ideas (see, for instance, \cite{resurgence1},
\cite{resurgence2}, \cite{[13]}, \cite{resurgence7}, \cite{resurgence8} and
references therein) with applications in Quantum Mechanics, topological
strings, and integrable Quantum Field Theory in low dimensions (see
\cite{[13]}, \cite{[14]}, \cite{[15]}, \cite{[16]}, \cite{[17]}, \cite{[6]},
\cite{Marino:2019wra} and references therein), however there are relatively
few applications of resurgence in non-integrable models. It is therefore
surprising that in the analysis of FFP interacting with Hadronic matter (which
is a sort of prototype of non-integrable systems), operators which can be
analyzed with such a tool (such as the Lam\'e operator) appear. In fact, one
dimensional periodic quantum systems like Mathieu and Lam\'e equations have
well known resurgence structures, and in the usual cases which one is the
suitable ``resurgent parameter" is pretty clear from the beginning
\cite{resurgence6}, \cite{resurgence9}, \cite{wkb}. On the other hand, in the
present situation in which gauged solitons at finite Baryon density generate a
FFP, there are many potential expansion parameters and, \textit{a priori},
it is not obvious which one is the ``most suitable" from the resurgent
viewpoint. We will see that our analysis shed light on this issue as well.

As a last remark, it is possible to disclose a relation of the present
(3+1)-dimensional solitons with the Gross-Neveu model and the Nambu-Jona
Lasinio model in (1+1)-dimensions (see \cite{thies0}, \cite{thies1},
\cite{thies2}, \cite{thies3}, \cite{thies4}, \cite{dunne1}, \cite{dunne2},
\cite{dunne3}).

This paper is organized as follows: In Section 2 the gauged NLSM is
introduced. In Section 3 we show that the gauged NLSM can be solved
analytically to obtain topological multi-solitons solutions. In Section 4 we
show that the electromagnetic field generated by the multi-solitons is a FFP,
and we study its main physical features, including the trajectory of charged
particles in this background. In Section 5, in order to test the stability of
the solutions, we perform perturbations on the solitons as well as on the
electromagnetic field. In Section 6 we draw a relation between the
perturbations on the solutions and the theory of resurgence. Section 7 is
devoted to the comparison of our solutions with the (1+1)-dimensional crystals
of the Gross-Neveu model. In the final section some conclusions will be drawn.


\section{The gauged non-linear sigma model}


The action of the gauged NLSM minimally coupled with the Maxwell theory is
\begin{gather}
I = \int d^{4}x\sqrt{-g}\left[  \frac{K}{4}\mathrm{Tr}\left(  L^{\mu}L_{\mu
}\right)  -m^{2}\mathrm{Tr}\left(  U+U^{-1}\right)  -\frac{1}{4}F_{\mu\nu
}F^{\mu\nu}\right]  \ ,\label{I}\\
L_{\mu} = U^{-1}D_{\mu}U \ , \quad\ F_{\mu\nu} = \partial_{\mu}A_{\nu
}-\partial_{\nu}A_{\mu} \ ,\label{I2}\\
U(x) \in SU(2)\ , \quad L_{\mu}=L_{\mu}^{j}t_{j}\ , \quad t_{j}=i\sigma_{j}
\ , \label{I3}%
\end{gather}
where $g$ is the determinant of the metric, $m$ is the Pions mass and $K$ is
the coupling constant of the NLSM. Here $A_{\mu}$ is the gauge potential,
$\nabla_{\mu}$ denotes the partial derivative and $\sigma_{i}$ are the Pauli
matrices. The covariant derivative is defined by
\[
\ D_{\mu}=\nabla_{\mu}+A_{\mu}\left[  t_{3},\ .\ \right]  \ .
\]
The field equations, obtained varying the action in Eq. \eqref{I} w.r.t the
fields $U$ and $A_{\mu}$, are
\begin{equation}
D_{\mu}L^{\mu}+\frac{2 m^{2}}{K} \left(  U-U^{-1}\right)  =0 \ , \label{NLSM}%
\end{equation}
\begin{equation}
\nabla_{\mu}F^{\mu\nu}=J^{\nu}\ , \label{maxwellNLSM}%
\end{equation}
where the current, $J^{\mu}$, in Eq. \eqref{maxwellNLSM} is given by
\begin{align}
J^{\mu}=\frac{K}{2}\text{Tr}\left[  \widehat{O}L^{\mu}\right]  \ ,\qquad
\widehat{O}=U^{-1}t_{3}U-t_{3}\ . \label{current}%
\end{align}
The energy-momentum tensor of the model is
\begin{equation}
T_{\mu\nu}=-\frac{K}{2}\mathrm{Tr}\left[  L_{\mu}L_{\nu}-\frac{1}{2}g_{\mu\nu
}L^{\alpha}L_{\alpha}\right]  -m^{2} \mathrm{Tr}\biggl[ g_{\mu\nu}(U+U^{-1})
\biggl] +\bar{T}_{\mu\nu}\ ,\nonumber
\end{equation}
with $\bar{T}_{\mu\nu}$ the electromagnetic energy-momentum tensor,
\begin{equation}
\bar{T}_{\mu\nu}=F_{\mu\alpha}F_{\nu}^{\;\alpha}-\frac{1}{4}F_{\alpha\beta
}F^{\alpha\beta}g_{\mu\nu}\ . \label{tmunu(1)}%
\end{equation}
On the other hand the topological charge is
\begin{equation}
B=\frac{1}{24\pi^{2}}\int_{\Sigma}\rho_{\text{B}}\ , \label{B}%
\end{equation}
where the topological charge density $\rho_{\text{B}}$ is defined by
\begin{equation}
\rho_{\text{B}}=\epsilon^{ijk}\text{Tr}\biggl[\left(  U^{-1}\partial
_{i}U\right)  \left(  U^{-1}\partial_{j}U\right)  \left(  U^{-1}\partial
_{k}U\right)  -\partial_{i}\left[  3A_{j}t_{3}\left(  U^{-1}\partial
_{k}U+\left(  \partial_{k}U\right)  U^{-1}\right)  \right]  \biggl]\ .
\label{B.1}%
\end{equation}
The above expression was constructed in \cite{Witten} (see also the
pedagogical analysis in \cite{gaugeI}), where the second term in Eq.
(\ref{B.1}) is added in order to guarantee both the conservation and the gauge
invariance of the topological charge. When $\Sigma$ is space-like, $B$ turns
out to be the Baryonic charge of the configuration.


\section{Gauged solitons at finite Baryon density}


In this section we show that the gauged $SU(2)$-NLSM admits analytic solutions
describing crystals of topological solitons.

\subsection{The ansatz}

As it has been already emphasized, the analytic description of cold and dense
nuclear matter as a function of the topological charge is a very interesting
but quite challenging theoretical problem (see \cite{R1}, \cite{R2} and
references therein). Since one of the main goals of the present analysis is
precisely to understand how gauged solitons react to the presence of a finite
volume containing them, we need to analyze the model in the following flat
metric%
\begin{equation}
\label{box}ds^{2}=-dt^{2}+L_{r}^{2}dr^{2}+L_{\theta}^{2}d\theta^{2}+L^{2}%
d\phi^{2} \ ,
\end{equation}
where the adimensional coordinates $\{r, \theta, \phi\}$ have the ranges
\begin{equation}
\label{ranges}0\leq r \leq2\pi\ , \quad0\leq\theta\leq\pi\ , \quad0\leq
\phi\leq2\pi\ .
\end{equation}
The above metric represents a box with different lengths on each side, so that
$V=4\pi^{3}L_{r}L_{\theta}L$, is the volume of the box in which the solitons
are confined.

For the $U$ field we adopt the standard parametrization of an element of
$SU(2)$, that is
\begin{equation}
\label{U}U^{\pm1}\left(  x^{\mu}\right)  =\cos(\alpha) \mathbf{1}_{2} \pm
\sin(\alpha) n^{i} t_{i} \ ,
\end{equation}
where $\mathbf{1}_{2}$ is the $2\times2$ identity matrix and%
\begin{gather}
n^{1} =\sin\Theta\cos\Phi, \quad n^{2}=\sin\Theta\sin\Phi, \quad n^{3}%
=\cos\Theta\ ,\label{n}\\
\alpha=\alpha(x^{\mu}) \ , \quad\Theta=\Theta(x^{\mu}) \ , \quad\Phi
=\Phi(x^{\mu}) \ , \quad n^{i} n_{i} = 1 \ .\nonumber
\end{gather}
According to Eqs. \eqref{B}, \eqref{B.1}, \eqref{U} and \eqref{n} it follows
that the topological charge density is
\begin{equation}
\label{rhoBexplicit}\rho_{\text{B}} = 12 (\sin^{2}\alpha\sin\Theta)
\ d\alpha\wedge d \Theta\wedge d \Phi\ .
\end{equation}
Here, from Eq. \eqref{rhoBexplicit} one can see that to have topologically
non-trivial configurations we must demand that $d\alpha\wedge d \Theta\wedge d
\Phi\ \neq\ 0 $ . On the other hand, as we want to construct analytical
solutions, it is necessary to have a good ansatz that significantly reduces
the field equations in Eqs. \eqref{NLSM} and \eqref{maxwellNLSM}. The approach
developed in \cite{crystal1}, \cite{crystal2} and \cite{CLV} lead to the
following ansatz for the gauged solitons
\begin{gather}
\alpha=\alpha(r) \ , \quad\Theta=q\theta\ , \quad\Phi=p\left(  \frac{t}%
{L}-\phi\right)  \ ,\label{hedgehog1}\\
q=2v+1 \ , \quad p,v\in\mathbb{N} \ , \quad p \neq0 \ ,\nonumber
\end{gather}
as well as for the electromagnetic potential
\begin{equation}
A_{\mu}=(u,0,0,-L u) \ , \qquad u=u(r,\theta) \ . \label{hedgehog2}%
\end{equation}
It is a direct computation to verify that the ansatz defined in Eqs.
(\ref{hedgehog1}) and (\ref{hedgehog2}) possesses the ``hedgehog property".
Such a property, which in the past was usually only associated to spherically
symmetric solitons, corresponds to the following desirable feature: one would
like to reduce the coupled system of non-linear field equations defining the
solitons to only one non-linear equation for the profile keeping alive the
topological charge. In the present case, quite remarkably, this property holds
despite the lack of spherical symmetry and despite the minimal coupling with
the Maxwell gauge field, as we will see below.

\subsection{Analytic crystal of topological solitons}

Using the ansatz in Eqs. \eqref{box}, \eqref{U}, \eqref{hedgehog1} and
\eqref{hedgehog2} one can check directly that, first of all, the three coupled
field equations of the gauged NLSM reduce to a single ODE for the profile
$\alpha$, namely
\begin{equation}
\alpha^{\prime\prime}-\frac{q^{2}}{2}\frac{L_{r}^{2}}{L_{\theta}^{2}}%
\sin(2\alpha)+\frac{4L_{r}^{2}m^{2}}{K}\sin(\alpha)\ =\ 0\ ,\label{Eqalpha}%
\end{equation}
which can be reduced to a first order equation of the form
\begin{equation}
\alpha^{\prime2}+\frac{q^{2}}{2}\frac{L_{r}^{2}}{L_{\theta}^{2}}\cos
(2\alpha)-\frac{8L_{r}^{2}m^{2}}{K}\cos(\alpha)\ =\ E_{0}\ ,\label{Eqalpha0}%
\end{equation}
with $E_{0}$ an integration constant. The above equation can be solved
explicitly in terms of Inverse Elliptic Functions and is reduced to the
following quadrature
\begin{equation}
dr=\pm\frac{1}{\eta(\alpha,E_{0})}d\alpha\ ,\quad\eta\left(  \alpha
,E_{0}\right)  =\pm\left[  2\left(  E_{0}-\frac{q^{2}}{4}\frac{L_{r}^{2}%
}{L_{\theta}^{2}}\cos(2\alpha)+\frac{4m^{2}L_{r}^{2}}{K}\cos(\alpha)\right)
\right]  ^{\frac{1}{2}}\ .\label{quadrature}%
\end{equation}
A necessary condition for stability is that $\alpha^{\prime}$ does not change
sign, therefore we must require, according to Eq. \eqref{quadrature}, that
\begin{equation}
E_{0}>\frac{q^{2}}{4}\frac{L_{r}^{2}}{L_{\theta}^{2}}+\frac{4m^{2}}{K}%
L_{r}^{2}\ .\label{E0constraint}%
\end{equation}
On the other hand, with the ansatz defined in Eqs. \eqref{box}, \eqref{U},
\eqref{hedgehog1} and \eqref{hedgehog2} the Maxwell equations are reduced to
just one linear Schr\"{o}dinger-like equation with an effective
two-dimensional potential, that is
\begin{equation}
\Delta\Psi+V\Psi\ =\ 0\ ,\qquad\Delta\equiv\frac{1}{L_{r}^{2}}\partial_{r}%
^{2}+\frac{1}{L_{\theta}^{2}}\partial_{\theta}^{2}\ ,\label{EqMax}%
\end{equation}
where we have defined
\begin{equation}
\Psi\ =\ \frac{2L}{p}u-1\ ,\qquad V=4K\sin^{2}(\alpha)\sin^{2}(q\theta
)\ .\label{V}%
\end{equation}
Note that the Maxwell equation in Eq. \eqref{EqMax} can be solved once the
soliton profile is determined, whose solution is implicit in Eq. \eqref{quadrature}.

\subsection{Topological charge and energy density}

From Eqs. \eqref{B}, \eqref{U}, \eqref{rhoBexplicit}, \eqref{hedgehog1} and
\eqref{hedgehog2} one can see that the natural boundary conditions in the
present case (see \cite{crystal2} for a detailed analysis) are the following
\begin{equation}
\label{bc}\alpha(2\pi)-\alpha(0)=n\pi\ , \qquad u(r,\pi)=(-1)^{q}u(r,0) \ .
\end{equation}
The above boundary conditions in Eq. \eqref{bc} determine that the topological
charge turns out to be
\begin{equation}
B=\left\{
\begin{array}
[c]{ll}%
-np & \text{ if }\quad q\in2\mathbb{Z}+1\\
0 & \text{ if }\quad q\in2\mathbb{Z}%
\end{array}
\right.  \ . \label{Bfinal}%
\end{equation}
Now, from Eqs. \eqref{ranges}, \eqref{quadrature} and \eqref{Bfinal} we see
that the integration constant $E_{0}$ is fixed in terms of $n$ through the
relation
\begin{equation}
n \int_{0}^{\pi} \frac{1}{\eta\left(  \alpha, E_{0}\right)  } d \alpha=2
\pi\ . \label{E0_relation_n}%
\end{equation}
The above equation always has a solution according to Eq.
\eqref{E0constraint}, which can be found numerically.

On the other hand, the energy density of the configurations presented above is
given by
\begin{align}
T_{00}\ =\  &  \frac{K}{2L_{r}^{2}L_{\theta}^{2}L^{2}}\biggl(L_{\theta}%
^{2}L^{2}\alpha^{\prime2}+2L_{r}^{2}\sin^{2}(\alpha)\biggl[L^{2}%
q^{2}+2L_{\theta}^{2}\sin^{2}(q\theta)(p-2Lu)^{2}\biggl]\biggl)\nonumber\\
&  +4m^{2}\cos(\alpha)+\frac{1}{L_{r}^{2}}(\partial_{r}u)^{2}+\frac
{1}{L_{\theta}^{2}}(\partial_{\theta}u)^{2}\ . \label{ED}%
\end{align}
It is also convenient to define the ``reduced energy density'' $\varepsilon
_{r}$ in the $r$-direction obtained integrating Eq. (\ref{ED}) along the
coordinates $\theta$ and $\phi$, that is
\begin{equation}
\varepsilon_{r}=\int\left(  L_{\theta}L\right)  T_{00}d\theta d\phi\ .
\label{ED1}%
\end{equation}
This reduced energy density represents the effective energy density (namely,
the energy per unit of length) in the $r$-direction, and can be compared
directly with 1-dimensional models possessing solitons crystals. In the
particular case in which both the Maxwell gauge potential and the Pions mass
vanish one gets
\begin{align}
\varepsilon_{r}^{(0)} \ =  &  \ \varepsilon_{r}(u=0,m=0)\nonumber\\
\ =  &  \ \frac{\pi K \sin^{2}(\alpha(r)) \left(  2 \pi q \left(  L^{2}
q^{2}+L_{\theta}^{2} p^{2}\right)  -L_{\theta}^{2} p^{2} \sin(2 \pi q)\right)
}{L L_{\theta} q}+\frac{\pi^{2} K L L_{\theta} \alpha^{\prime2}}{L_{r}^{2}}
\ . \label{ED2}%
\end{align}


\section{On the force free nature of the electromagnetic field}


In this section we will show that the electromagnetic field generated by the
gauged solitons discussed in the previous sections satisfies the FFP
condition. To the best of the authors knowledge, this is the first analytic
example in which a FFP emerge in a natural way.

\subsection{Emergent force free plasma}

The force free condition
\begin{equation}
F_{\mu\nu}\nabla_{\rho}F^{\nu\rho}=0\ , \label{ffc2}%
\end{equation}
is realized in many relevant examples of plasmas such as in the solar corona
\cite{WS2012} or close to rotating neutron stars and black holes
\cite{Gold1968}, \cite{Pacini1968}, \cite{GW1969}. As it was already mentioned
the FFP plays an important role in the Blandford-Znajek process \cite{BZ} but,
even in such process, the FFP emerge as a suitable choice that ``provide a
reasonable approximation to the time-averaged structure of the magnetosphere",
and then, the black hole parameters are estimate values in order to satisfy
the force-free condition.

It can be readily seen that the Maxwell field surrounding our gauged solitons
satisfies the force-free condition in Eq. \eqref{ffc2}. First, note that the
$U(1)$ gauge potential in Eq. \eqref{hedgehog2} can be written as
\[
A_{\mu}dx^{\mu}=u(r,\theta)\left(  dt-Ld\phi\right)  \ ,
\]
so that the field strength is found to be
\[
F_{\mu\nu}=\left(  \left(  \partial_{r}u\right)  dr+\left(  \partial_{\theta
}u\right)  d\theta\right)  \wedge\left(  dt-Ld\phi\right)  \ .
\]
It follows that the divergence of the field strength, identified with the
current, is
\begin{gather*}
J^{\nu} =\nabla_{\mu}F^{\mu\nu}=c_{1}\sin^{2}(\alpha)\sin^{2}(q\theta)\left(
\partial^{\nu}\Phi-c_{2}A^{\nu}\right)  \ ,\\
\text{with} \qquad c_{1} = -\frac{2K}{L},\ \ c_{2}=2L\ ,
\end{gather*}
which is manifestly orthogonal to the field strength $F_{\mu\nu}$. Therefore
we see that the force-free condition in Eq. \eqref{ffc2} is automatically satisfied.

The electromagnetic field lines (for suitable choice of the parameters) are
presented in Fig. \ref{fig1:image1}, where we have used the following boundary
conditions for the Maxwell potential
\begin{equation}
u(r, 0)=u(r, \pi) = 0 \ , \quad u(0, r)=u(2\pi, 0) = 0 \ .
\end{equation}
\begin{figure}[th]
\centering
\includegraphics[scale=0.25]{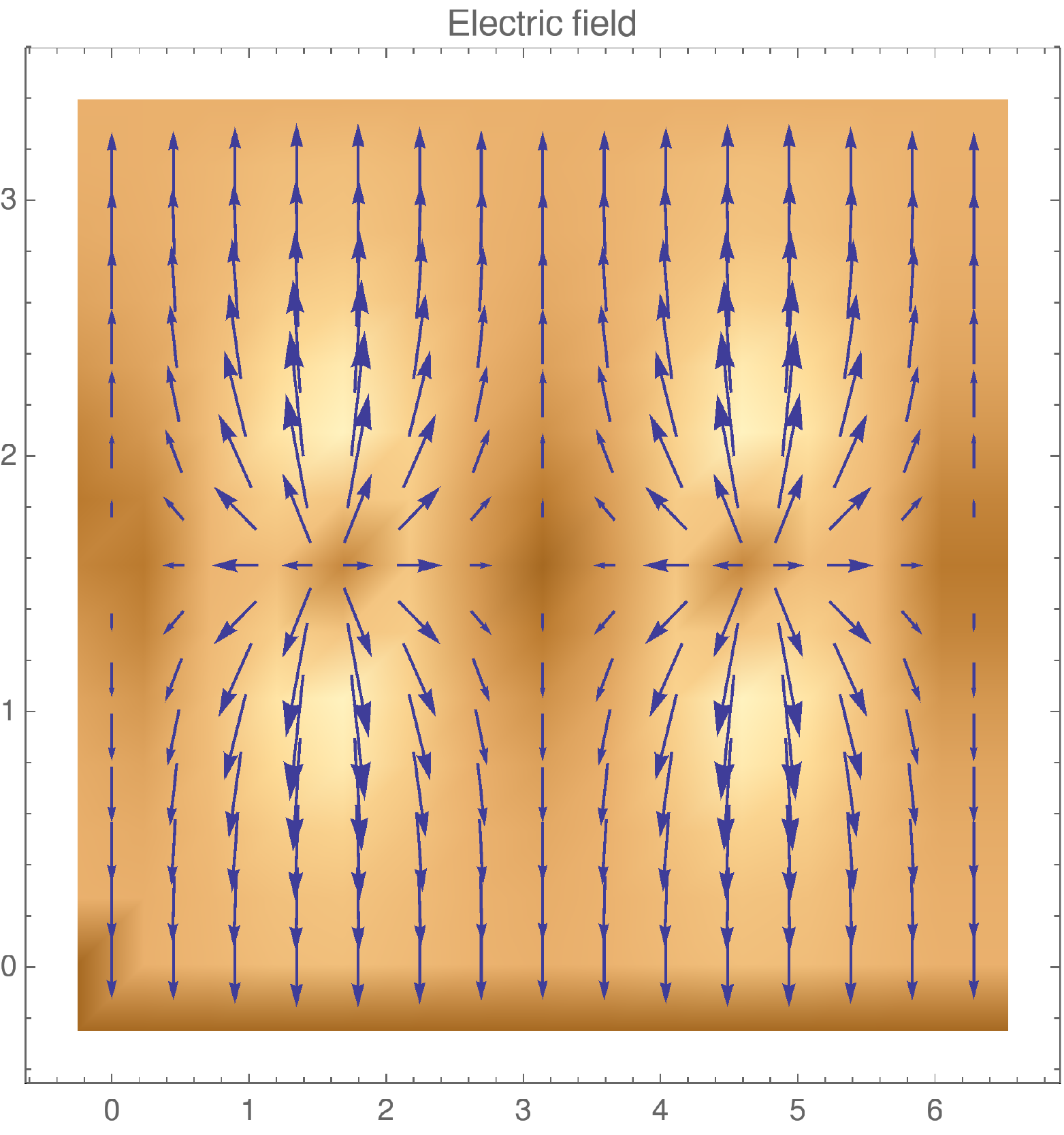}\quad
\includegraphics[scale=0.35]{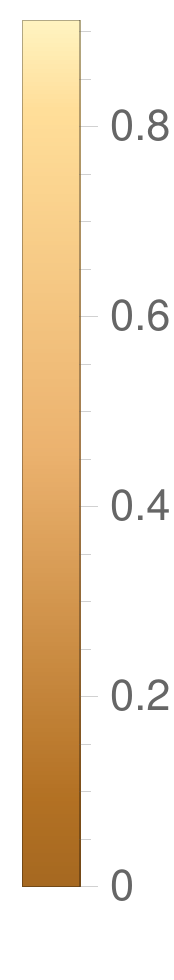}\quad
\includegraphics[scale=0.25]{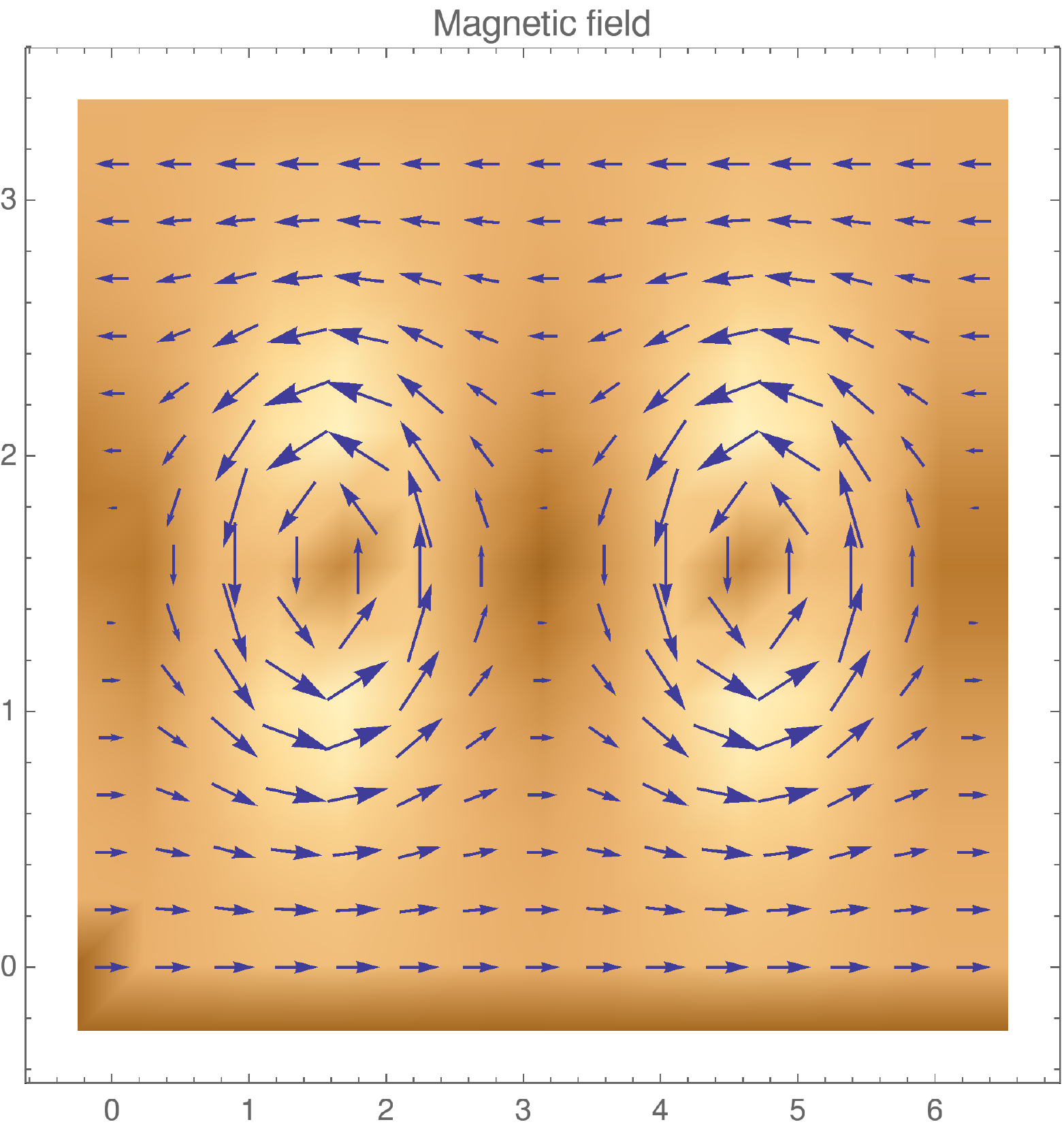}\quad
\includegraphics[scale=0.35]{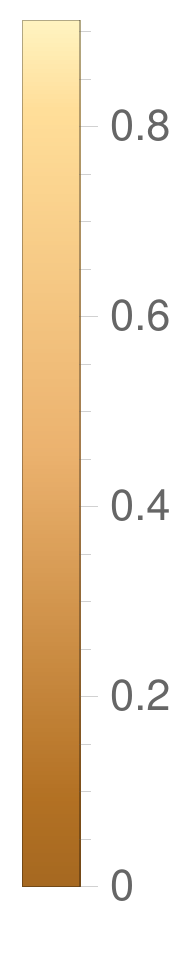}\quad
\includegraphics[scale=0.25]{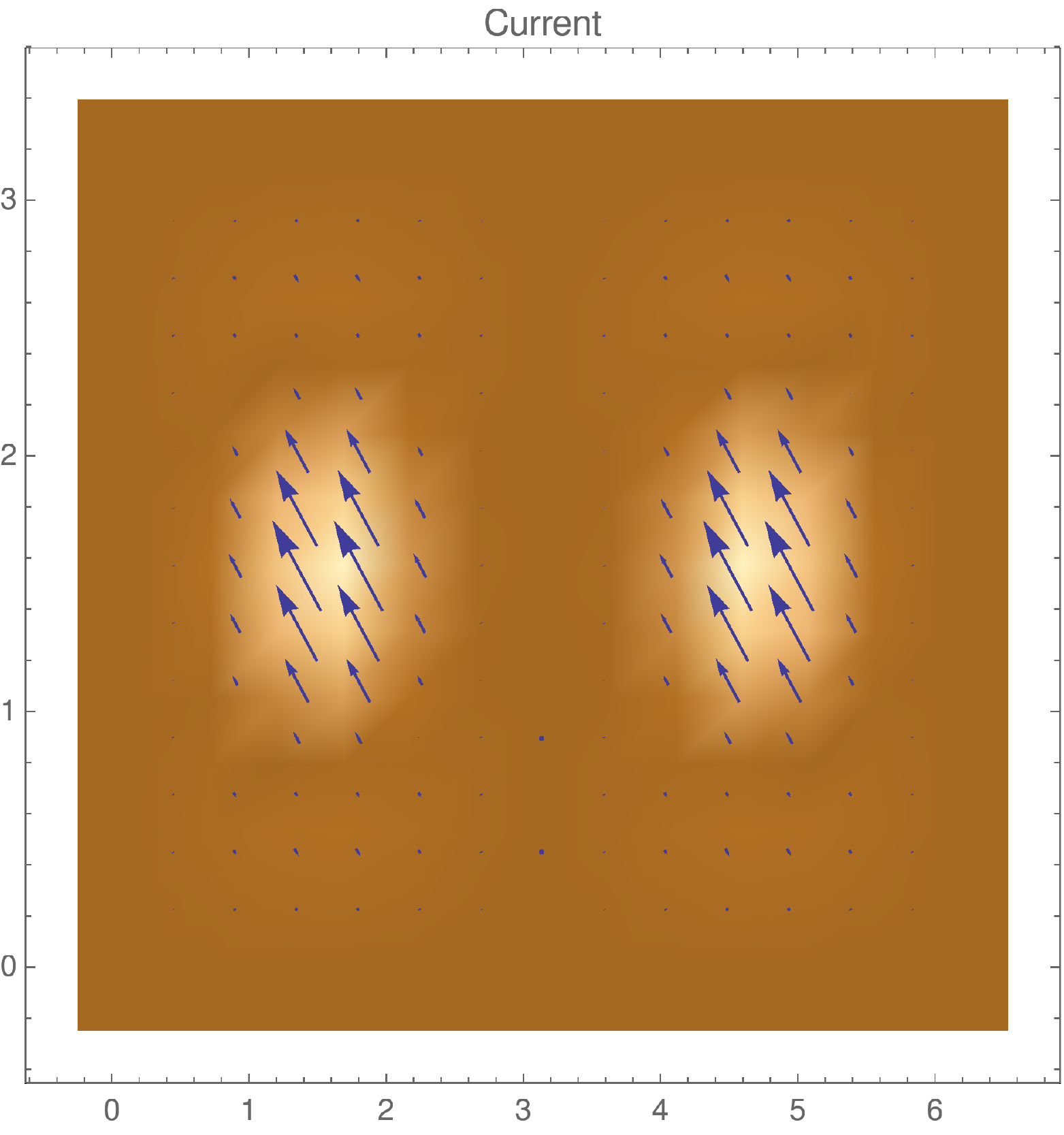}\quad
\includegraphics[scale=0.35]{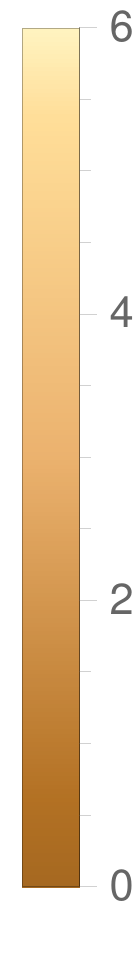} \caption{Electric field, magnetic
field and the current of two gauged solitons, with $n=2$, $m=0$, and $p=q=1$.
The electric and magnetic fields vanish in the center of the tubes while the
current is completely contained inside these.}%
\label{EM}%
\end{figure}
\begin{figure}[th]
\centering
\includegraphics[scale=0.4]{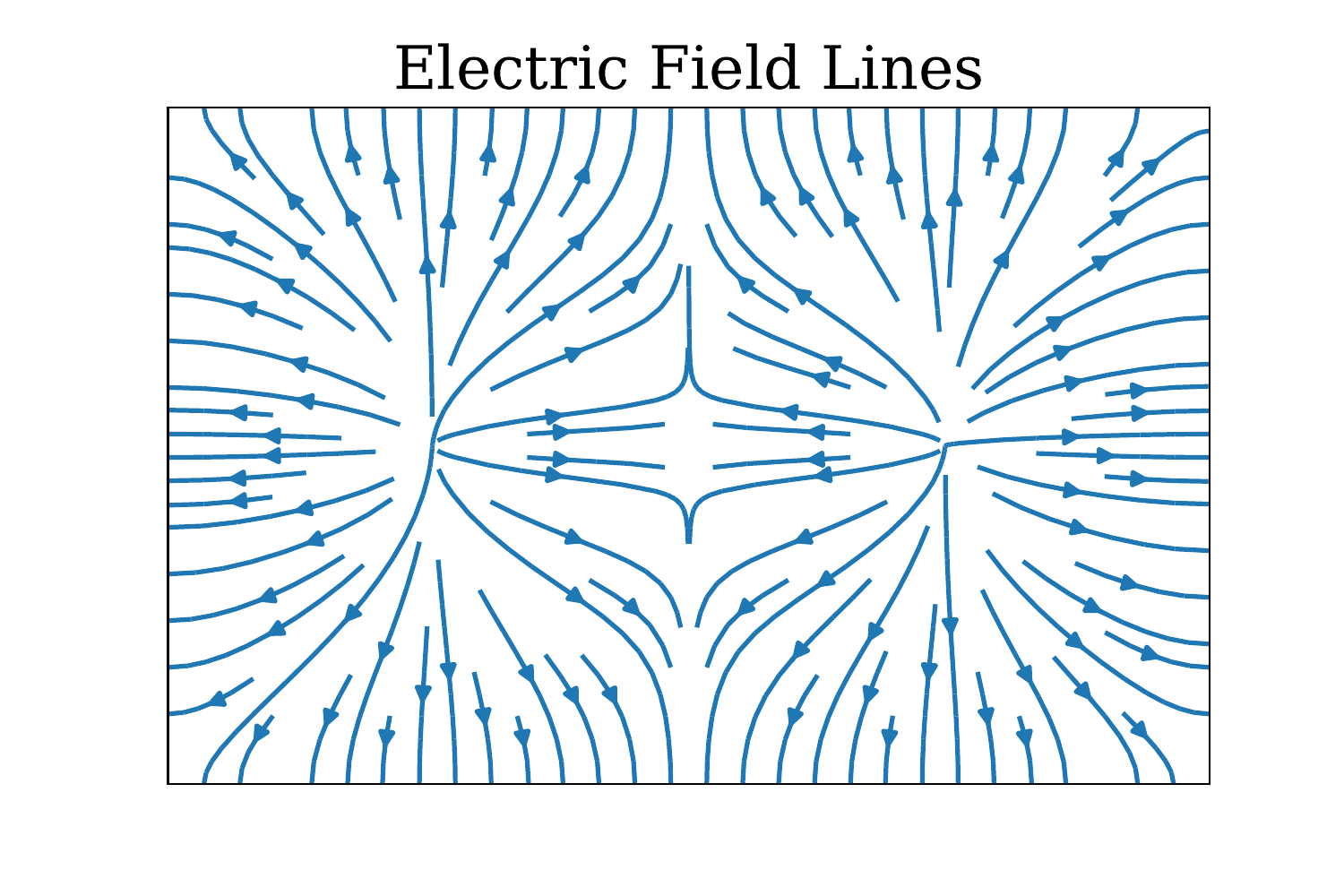} \qquad
\includegraphics[scale=0.4]{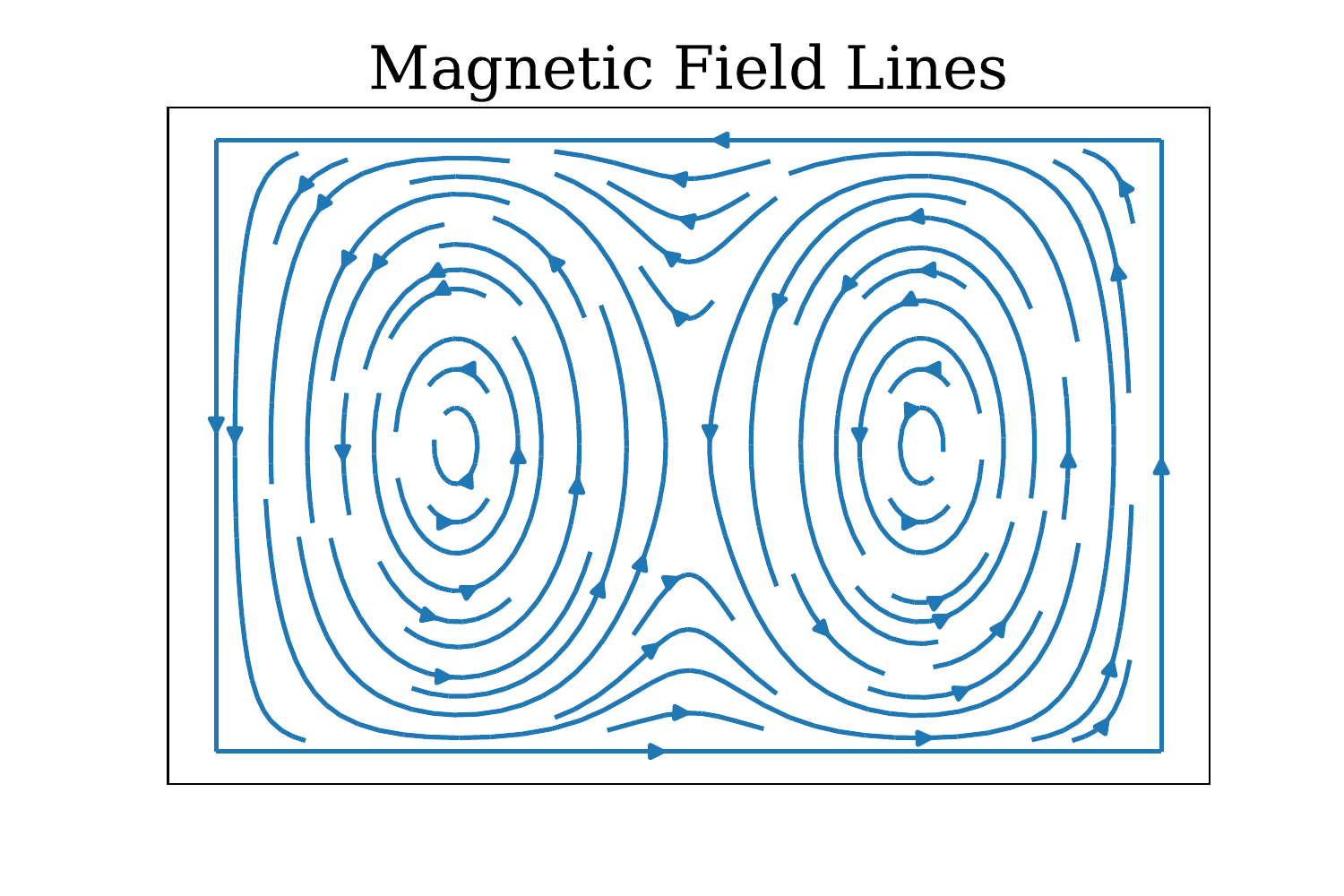}
\caption{Electric field lines and magnetic field lines of two gauged
solitons, with $n=2, m=0$, and $p=q=1$. }%
\label{fig1:image1}%
\end{figure}
At this point it is important to emphasize that, although several
analytic solutions to the force-free Maxwell equations could be found (for
instance, using the approaches in \cite{BGJ}, \cite{BG}, \cite{YZ},
\cite{ZYL}, \cite{ZMP}, \cite{GJ2} and \cite{CO}) most of these analytic
solutions to the force-free electrodynamics do not discuss the corresponding
sources. Instead, the source of the FFP electromagnetic field found here can
be clearly identified with the Hadronic degrees of freedom described by the NLSM.

\subsection{Trajectory of charged particles on the FFP}

Here we will draw some qualitative plots of the trajectories of charged test
particles (like electrons) on the FFP generated by the gauged solitons.

First of all, it follows from Eq. \eqref{hedgehog2} that the electric and
magnetic fields generated by the gauged solitons are
\begin{equation}
\vec{E}=(E_{r}, E_{\theta}, 0 ) \ , \qquad\vec{B}=(B_{r}, B_{\theta}, 0 ) \ ,
\end{equation}
where the components read
\begin{equation}
E_{r}=-\partial_{r} u \ , \quad E_{\theta}=-\partial_{\theta}u \ , \quad
B_{r}= \frac{1}{L^{3}}\partial_{\theta}u \ , \quad B_{\theta}=-\frac{1}{L^{3}%
}\partial_{r} u \ .
\end{equation}
For a test particle of charge $q_{e}$ and velocity $\vec{v}$, the Lorentz
force generated by the FFP acting on the particle is
\begin{equation}
\label{Lorentz}\vec{F} \ = \ q_{e}(\vec{E} + \vec{v}\times\vec{B}) \ ,
\end{equation}
with $\vec{v}=(v_{r},v_{\theta},v_{\phi})$.

\begin{figure}
\centering
\includegraphics[scale=0.35]{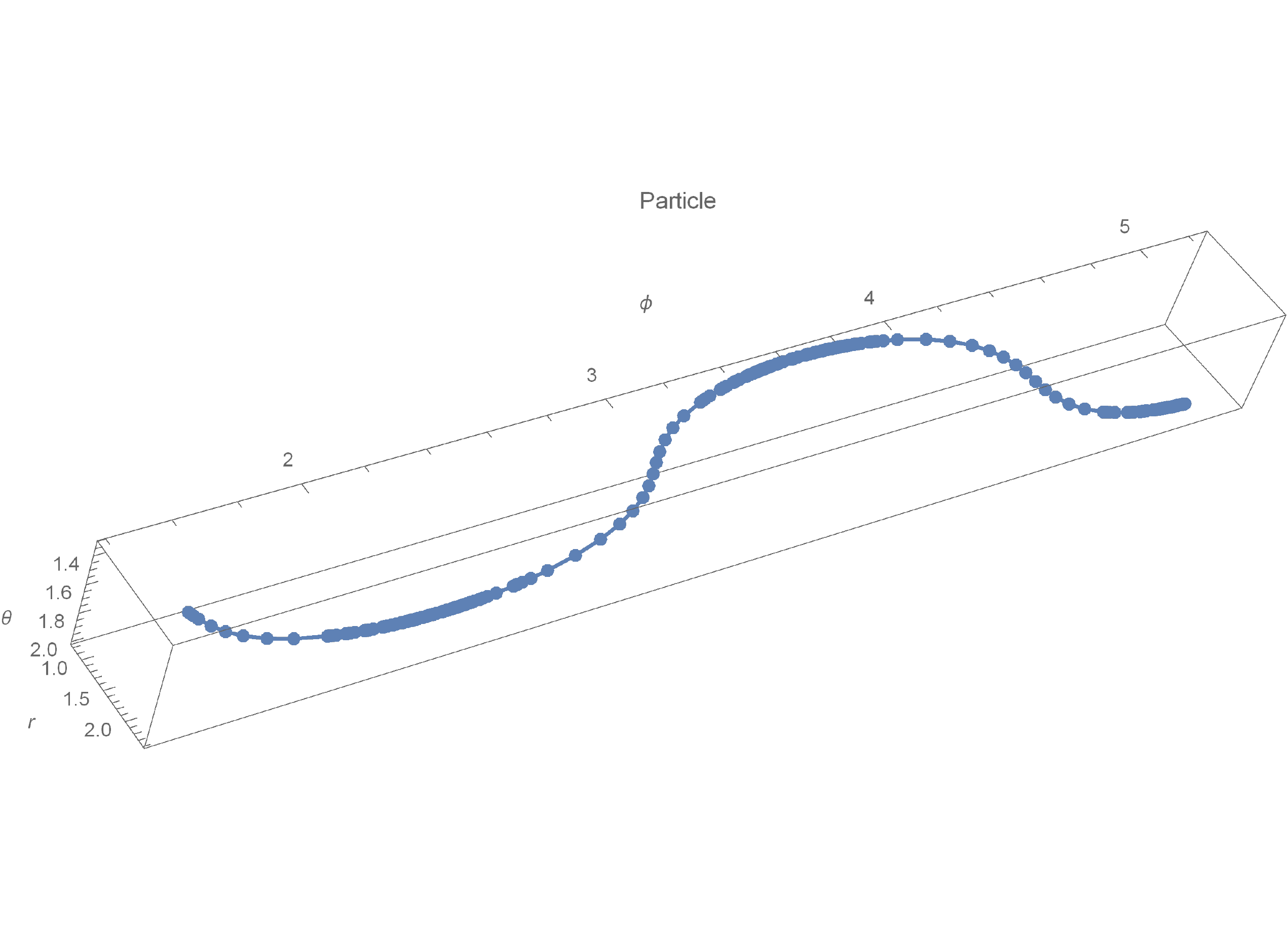}
\includegraphics[scale=0.35]{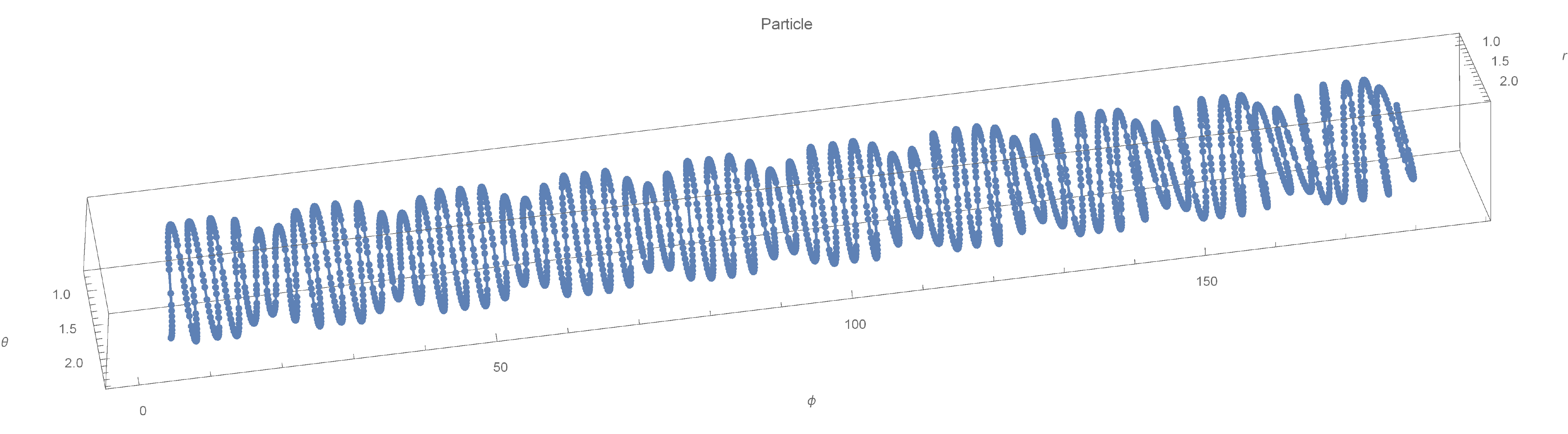}
\includegraphics[scale=0.35]{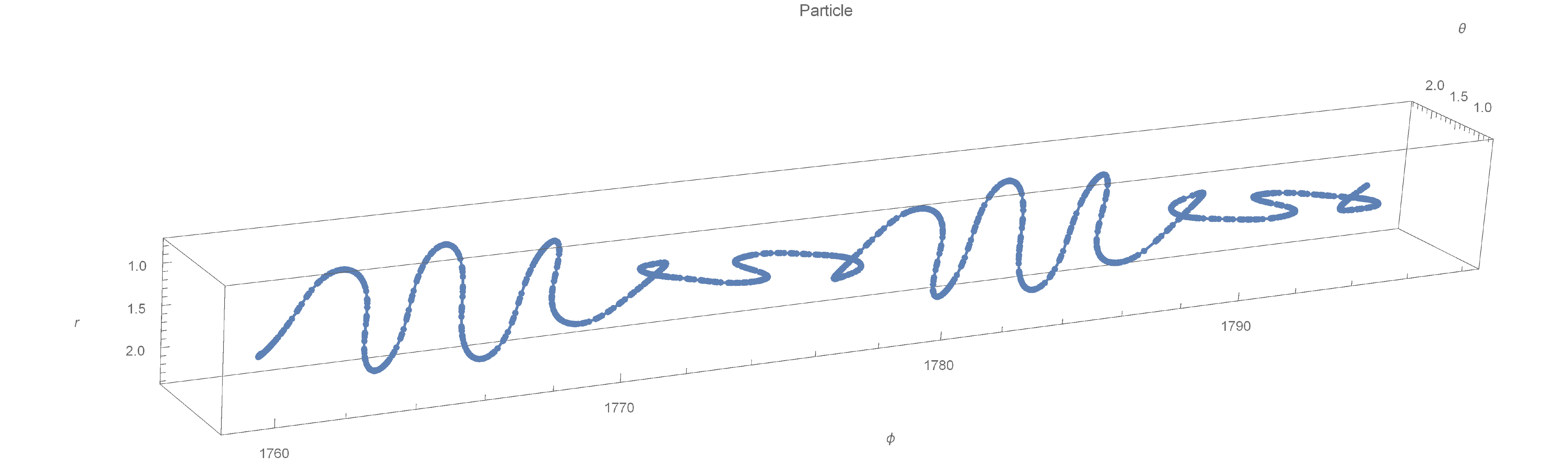}
\caption{Trajectory of a single
particle under the Lorentz force in a time interval between 0 and 1 in (a), a time interval between 1 and 50 in (b) and a time interval between 490 and 500 in (c). The arrow denotes the ``increasing time direction" on the trajectory of the test particle. The dots on the trajectory have been taken at equal time interval. Thus, regions where neighbouring dots are ``far apart" correspond to ``high velocity" regions of the trajectory.}%
\label{alllorentz}%
\end{figure}

Eq. \eqref{Lorentz} is a set of three coupled differential equations, namely
\begin{align}
\frac{m}{q_{e}}\frac{d^{2}r}{dt^{2}}\ = &  \ E_{r}-B_{\theta}\frac{d\phi}%
{dt}\ ,\\
\frac{m}{q_{e}}\frac{d^{2}\theta}{dt^{2}}\ = &  \ E_{\theta}+B_{r}\frac{d\phi
}{dt}\ ,\\
\frac{m}{q_{e}}\frac{d^{2}\phi}{dt^{2}}\ = &  \ B_{\theta}\frac{dr}{dt}%
-B_{r}\frac{d\theta}{dt}\ .
\end{align}
The above equations can be numerically integrated to know the trajectory of
test particles. We show the trajectory of a single particle for times $0<t<1$, for $1<t<50$ and for $490<t<500$
in Fig. \ref{alllorentz}.
In these figures one can notice that the charged test particle oscillates in
the $r-\theta$ plane while moving along the axis of the tube. It is also worth
to emphasize that this motion is quite different from the usual trajectory of
a charged test particle in the magnetic field of a thin straight wire. Indeed,
in the case of a thin solenoid there is only the $\theta-$component of the
magnetic field (and no electric field) while in the present case there is also
a radial component and both the electric and the magnetic fields are non-vanishing.


\section{Perturbative analysis}


A complete perturbative analysis of these solutions and the corresponding FFP
is beyond the scope of the present paper (since such an analysis would involve
seven coupled PDE in the non-trivial background provided by the soliton
crystal itself). This task is very hard even from a numerical viewpoint.
Hence, in this section, we will analyze two special types of perturbations of
the Hadronic field and of the FFP which have both interesting physical meaning
and allows some analytic control.

\subsection{$SU(2)$ perturbations}

A relevant perturbation of these gauged solitons is the one which keeps the
hedgehog structure intact, namely, which keep alive the hedgehog property
defined in the previous section. In the present context such perturbations are
defined by the following small deformations of the original ansatz in Eqs.
\eqref{U} and \eqref{hedgehog1}:
\begin{equation}
\alpha(r)\rightarrow\alpha(r)+\varepsilon F_{1}(r)\ ,\qquad\Phi(t,\phi
)\rightarrow\Phi(t,\phi)+\varepsilon F_{2}(r)\ ,\label{pertype1}%
\end{equation}
with $\ 0<\varepsilon\ll1$. 
Such perturbations (and, in particular, the little disturbance $F_{1}$ of the
profile $\alpha(r)$) are very likely to be the \textquotedblleft smallest
energy perturbations"\footnote{The reason is that it takes \textquotedblleft a
little effort" to perform a radial deformation of $\alpha$ (see, for instance,
the discussion of the \textquotedblleft hedgehog ansatz" in \cite{shifman1}%
\ and \cite{shifman2}) as compared, for instance, to deformations of the
Isospin degrees of freedom $\Theta$ and $\Phi$. For instance, the classic
reference \cite{ANW} showed that the low energy \textquotedblleft Isospin
perturbations" have a gap. Hence, there are examples in the literatures in
which the unstable negative energy-modes are precisely in this sector while
the perturbations of the Isospin degrees of freedom have positive energies.}
of the Hadronic degrees of freedom. For this reason the linear operator which
determines the spectrum of these perturbations has a very important role.

The field equations in Eq. \eqref{NLSM} at first order in $\varepsilon$ for
the perturbations defined in Eq. (\ref{pertype1}) read%
\begin{equation}
\wp_{(1)}F_{1}=0\ ,\qquad\wp_{(2)}F_{2}=0\ ,\label{F1yF2}%
\end{equation}
where we have defined
\begin{align*}
\wp_{(1)} &  =-\frac{d^{2}}{dr^{2}}+\frac{L_{r}^{2}}{KL_{\theta}^{2}}\left(
Kq^{2}\cos(2\alpha)-4L_{\theta}^{2}m^{2}\cos\alpha\right)  \ ,\\
\wp_{(2)} &  =-\frac{d}{dr^{2}}-\cot(\alpha)\alpha^{\prime}\ .
\end{align*}
First of all, one can notice that $F_{2}(r)$ can be readily integrated giving
rise to%
\begin{equation}
F_{2}(r)=\int_{1}^{r}c_{1}\csc^{2}(\alpha(\gamma))d\gamma+c_{2}\ .
\end{equation}
On the other hand the situation for $F_{1}(r)$ is different. We will write
down such linear operator in the case $m=0$ since in this way the analysis is
cleaner, however a non-vanishing $m$ will not change the results and will only
make the analytic formulas a bit more involved. In the $m=0$ case the soliton
profile (see Eqs. \eqref{Eqalpha} and \eqref{Eqalpha0}) is given by
\begin{equation}
\alpha^{\prime}=\pm\left[  2\left(  E_{0}-\frac{q^{2}}{4}\frac{L_{r}^{2}%
}{L_{\theta}^{2}}\cos(2\alpha)\right)  \right]  ^{\frac{1}{2}}\ ,\quad
\alpha(0)=0\ ,\label{pertex1}%
\end{equation}%
\begin{equation}
\alpha(r)=\pm\operatorname{am}\left(  \sqrt{-\frac{L_{r}^{2}q^{2}}{L_{\theta
}^{2}}}\frac{r}{k};k\right)  \ ,\quad k=\sqrt{\frac{2L_{r}^{2}q^{2}}{L_{r}%
^{2}q^{2}-4E_{0}L_{\theta}^{2}}}\ ,\label{pertex2}%
\end{equation}
where $\operatorname{am}(x;k)$ is the the Jacobi Amplitude \cite{math} and $k$
is the elliptic parameter (related to the \textit{elliptic modulus $m$} in
\cite{math} as $k\equiv\sqrt{m}$). From here on we are going to use the
parameter $k$ instead $m$. The periodic structure appears due to the
well-known properties of the function $\operatorname{am}(r;k)$, namely
\begin{align}
\sin{(\operatorname{am}(r;k))} &  =\operatorname{sn}(r;k)\ ,\label{snam}\\
\cos{(\operatorname{am}(r;k))} &  =\operatorname{cn}(r;k)\ ,\label{cnam}\\
\sqrt{1-k^{2}\sin^{2}(\mathrm{am}(u,k))} &  =\operatorname{dn}%
(u,k)\ ,\label{dnam}%
\end{align}
where $\operatorname{sn}(r;k)$, $\operatorname{cn}(r;k)$ and
$\operatorname{dn}(r;k)$ are the Jacobi Elliptic Functions \cite{math}.The
function $\operatorname{sn}^{2}(r;k)$ is periodic, with period $2K(k)$, where
$K(k)=\int_{0}^{\pi/2}d\theta/\sqrt{1-k\sin^{2}\theta}$ is the
elliptic-quarter period.

Going back to the linearized equations, the equation for $F_{1}(r)$ in Eq.
\eqref{F1yF2} in the case $m=0$ takes the form
\begin{equation}
\wp_{(0)} F_{1}=0 \ , \qquad\wp_{(0)} =-\frac{d^{2}}{dr^{2}}+\frac{L_{r}%
^{2}q^{2}}{L_{\theta}^{2}} \cos(2\alpha)\ .
\end{equation}
Quite interestingly, using the properties of Jacobi Amplitudes in Eqs. \eqref{snam}, \eqref{cnam} and \eqref{dnam}, the operator $\wp_{(0)}$ (which
determines the stability of the present gauged solitons system under the
perturbations in Eq. \eqref{pertype1}) can be cast in
the form of a Lam\'e Operator:%
\begin{align}
\wp_{(0)}  &  =-\frac{d^{2}}{dr^{2}}+\frac{L_{r}^{2}q^{2}}{L_{\theta}^{2}%
}(\operatorname{cn}(r/k;k)^{2}-\operatorname{sn}(r/k;k)^{2})\ . \label{lame2}%
\end{align}
Consequently, the spectrum of the family of operators $\wp_{(0)}$ (or
$\wp_{(m)}$) defined above is very important not only in relation with the
stability analysis of the present gauged solitons but also because such
spectrum encodes relevant physical informations about the band spectrum of
these gauged solitons.

As far as the stability of this gauged solitons-FFP system one need to study
the following eigenvalues problem
\begin{equation}
\wp_{(0)}\Psi=\Omega^{2}\Psi\ ,\label{constab1}%
\end{equation}
so that the stability condition under this family of perturbations is%
\begin{equation}
\Omega^{2}\geq0\ .\label{constab2}%
\end{equation}
It can be seen that the condition in Eq. (\ref{constab2}) can be satisfied.
Indeed, it is easy to find a zero mode just taking%
\[
\Psi_{0}=\partial_{r}\alpha_{0}\qquad\Rightarrow\quad\wp_{(0)}\Psi_{0}=0\ ,
\]
where $\alpha_{0}$ is a background solution of Eqs. \eqref{pertex1} and
\eqref{pertex2} (or Eq. \eqref{Eqalpha} in the case with $m\neq0$). Moreover
we can also deduce that $\Psi_{0}$ has no node since $\partial_{r}\alpha_{0}$
does not vanish. Consequently, standard theorems in Quantum Mechanics ensure
that $\Omega^{2}=0$ is the minimal eigenvalue and all the other are positive.
\ In fact, Eq. (\ref{constab1}) encodes many more information since it has to
do with the spectrum of the lowest energy perturbations of the present gauged
solitons. In other words, the spectrum of $\wp_{(0)}$ (or $\wp_{(m)}$) will
tell us the \textquotedblleft phonons" of the system. In the next section we
will analyze such spectrum in the case $m=0$ in which the results are cleaner
and directly related with the theory of resurgence. The results with $m\neq0$
are very similar but more difficult to analyze with resurgence since such
spectrum involves two relevant parameters (this case belongs to the so-called
parametric resurgence).

Last but not least, this family of operators is especially well-suited for the
resurgence analysis as it has been shown in \cite{[17]}, \cite{resurgence6}
and \cite{resurgence9}. One of the main results of this section is that the
proper resurgent parameter $g$ (in the $m=0$ case) is%
\begin{equation}
\frac{1}{g^{2}} = \frac{L_{r}^{2}q^{2}}{L_{\theta}^{2}} \ . \label{resurgent1}%
\end{equation}
Although this result is very simple (one just needs to compare Eq.
\eqref{lame2} with those in \cite{[17]}, \cite{resurgence6} and
\cite{resurgence9}) it is very non-trivial. It is relevant to note that the
effective resurgent parameter\footnote{One of the requirements to be a ``good
resurgent parameter" is that, for instance, it should determine the spectrum
of the small fluctuation of the profile. In particular, when moving the
parameter from small to large, one should get the transition in the spectrum
from small gaps to large gaps as in the Mathieu equation \cite{[17]}.} $g$
does not involve the coupling constant of the theory, namely $K$. Instead, the
direct computation discussed in this section shows that the suitable parameter
is in Eq. (\ref{resurgent1}), which depends on the odd integer $q$ as well as
on the ``asymmetry ratio" $\frac{L_{r}^{2}}{L_{\theta}^{2}}$\ which defines
how far from a square is the basis in the $r-\theta$ plane of the box in which
these gauged solitons are living. The reason why we were able to derive this
result explicitly is ``just" that we have constructed analytically these
gauged solitons which are sources of FFP. Without analytical solutions it
would be impossible a proper identification of the correct resurgent parameter.

It is worth to emphasize the similarity of the spectrum of the ``phonons"
defined by the operator in Eq. (\ref{constab1}) with the ones of the crystal
of kinks in \cite{SGkink} as well as in the analysis of the small fluctuations
of solitons crystals in integrable field theory models in (1+1) dimensions. In
Section 7 we will elaborate more on this comparison.

\subsection{Electromagnetic perturbations}

An effective technique to analyze the stability of solutions presented here is
to consider the gauged solitons as an effective background medium on which the
electromagnetic perturbations propagate. This is an excellent approximation
especially in the 't Hooft limit because in the semiclassical Photon-Baryon
scattering, the Baryon is basically not affected since the Photon has zero
mass (for a detailed discussion see \cite{5b}). Consequently, the Photons
perceive the gauged soliton as an effective medium while the solitons are not
affected by the small fluctuations of the electromagnetic field.
Correspondingly, in this section we will consider the $SU(2)$ degrees of
freedom as fixed to be the background solution. As it has been already
emphasized, the complete stability analysis not only is completely out of
reach from analytical methods but also numerically is a very hard task.
Nevertheless, the perturbations we are considering here encode relevant
features of the gauged solitons and of the corresponding FFP, as we will see below.

Let us consider the following perturbations on the Maxwell potential
\begin{equation}
(u,0,0,-Lu)\rightarrow(u+\varepsilon\xi_{1}(x^{\mu}),0,0,-Lu+\varepsilon
\xi_{2}(x^{\mu}))\ ,\label{pertu}%
\end{equation}
with $0<\varepsilon\ll1$. At first order in the parameter $\varepsilon$ and
for the perturbation defined in Eq. \eqref{pertu} the Maxwell equations in Eq.
\eqref{maxwellNLSM} become
\begin{gather*}
\partial_{\theta}(\partial_{\phi}\xi_{2}-L^{2}\partial_{t}\xi_{1})\ =\ 0\ ,\\
\partial_{r}(\partial_{\phi}\xi_{2}-L^{2}\partial_{t}\xi_{1})\ =\ 0\ ,\\
\biggl(\frac{1}{L_{r}^{2}}\partial_{r}^{2}+\frac{1}{L_{\theta}^{2}}%
\partial_{\theta}^{2}+\frac{1}{L^{2}}\partial_{\phi}^{2}\biggl)\xi_{1}%
-\frac{1}{L^{2}}\partial_{\phi}\partial_{t}\xi_{2}+4K\sin^{2}(\alpha)\sin
^{2}(q\theta)\xi_{1}\ =\ 0\ ,\\
\biggl(\frac{1}{L_{r}^{2}}\partial_{r}^{2}+\frac{1}{L_{\theta}^{2}}%
\partial_{\theta}^{2}-\partial_{t}^{2}\biggl)\xi_{2}+\partial_{\phi}%
\partial_{t}\xi_{1}+4K\sin^{2}(\alpha)\sin^{2}(q\theta)\xi_{2}\ =\ 0\ .
\end{gather*}
Since we want to test linear stability we need to check the (absence of)
growing modes in time. This implies that $\xi_{1}$ and $\xi_{2}$ must depend
on the temporal coordinate. But, according to the previous equations if
$\xi_{1}$ and $\xi_{2}$ depend on time these functions must also depend on the
coordinate $\phi$, that is
\begin{equation}
\partial_{t}\xi_{i}\neq0\quad\Rightarrow\partial_{\phi}\xi_{i}\neq
0\ ,\qquad\xi_{i}=\{\xi_{1},\xi_{2}\}\ .\label{electropert1}%
\end{equation}
For simplicity we will assume that
\begin{equation}
\partial_{\phi}\xi_{2}=L^{2}\partial_{t}\xi_{1},\label{electropert2}%
\end{equation}
so that the above equations system is reduced to
\[
\Box\xi_{i}+V\xi_{i}\ =\ 0\ ,\qquad\Box\equiv-\partial_{t}^{2}+\frac{1}%
{L_{r}^{2}}\partial_{r}^{2}+\frac{1}{L_{\theta}^{2}}\partial_{\theta}%
^{2}+\frac{1}{L^{2}}\partial_{\phi}^{2}\ ,
\]
with $V$ defined in Eq. \eqref{V}. Then, performing the Fourier transformation
in the coordinate $\phi$ and time $t$,
\[
\xi_{i}(t,r,\theta,\phi)\ =\ \int\widehat{\xi}_{i}(\omega,r,\theta
,k_{3})e^{-i(k_{3}\phi+\omega t)}dk_{3}d\omega\ ,
\]
we obtain an eigenvalue equation for $\hat{\xi}_{i}$,%
\begin{equation}
-\Delta\widehat{\xi}_{i}+V_{\text{eff}}\widehat{\xi}_{i}\ =\ \omega
^{2}\widehat{\xi}_{i}\ ,\label{eigenvalue_2d}%
\end{equation}
with
\begin{equation}
\Delta=\frac{1}{L_{r}^{2}}\partial_{r}^{2}+\frac{1}{L_{\theta}^{2}}%
\partial_{\theta}^{2}\ , \qquad V_{\text{eff}}=\frac{\left(  k_{3}\right)
^{2}}{L^{2}}-V\ , \qquad k_{3}\neq0\ . \label{Veff}%
\end{equation}
The non-vanishing eigenvalue
\begin{equation}
k_{3}=\frac{l}{(2\pi)}\ \label{Veff2} \ , 
\end{equation}
in the effective potential in Eq. \eqref{Veff} is the wave-number along the
$\phi$-direction, with $l$ a non-vanishing integer. A sufficient condition
ensuring linear stability under the perturbation defined in Eqs.
(\ref{electropert1}) and (\ref{electropert2}) is the requirement
\begin{equation}
V_{\text{eff}}>0\ . \label{requstab1}%
\end{equation}
The obvious reason is
that the above condition implies that the eigenvalues of the operator
$\widehat{O}_{S}$ defined as
\[
\widehat{O}_{S}=-\Delta+V_{\text{eff}} \ ,
\]
are positive and, consequently, the parameter $\omega$ in the time-Fourier transform of
the perturbations is real (so that there are no growing modes in time).
 
However, this condition is not necessary in the sense that it could be
possible for $V_{\text{eff}}$ to be ``slightly negative" keeping, at the same
time, the eigenvalue $\omega^{2}$ in Eq. (\ref{eigenvalue_2d}) real and
positive. On the other hand, the mathematical task to find a sharp
characterization of the stability in this sector can be very complicated from
the viewpoint of functional analysis in the case of Schrodinger-like
potentials which depend in a non-trivial way on two (or more) spatial
coordinates. Hence, here we will only consider the criterion in Eq.
(\ref{requstab1}) which is more than enough to provide a qualitative picture.
We hope to come back on this interesting mathematical issue in a future publication.

The requirement in Eq. (\ref{requstab1}) imposes an upper bound for the length
of the box. This can be seen as follows. The ``less favorable case" (from the
stability viewpoint) corresponds to the least possible value for $\left(
k_{3}\right)  ^{2}$ in Eq. (\ref{Veff2}) when the positive part of the
effective potential is the smallest. Correspondingly, Eq. (\ref{requstab1})
becomes
\begin{equation*}
\frac{\left(  k_{3}\right)  ^{2}}{L^{2}}-4K>0 \qquad \Rightarrow  
\quad  L<\frac{1}{4\pi\sqrt{K}}\approx1\ fm\ .
\end{equation*}
In terms of the Baryon density, the above bound implies that the present
configurations are viable when the Baryon density is of the order of 1 Baryon
per $fm^{3}$ or higher (this range is well within the range of validity of the
NLSM). In fact, we expect that a more refined mathematical analysis would give
a better bound showing that these solutions are viable even at lower densities.


\section{Some comments on the fluctuations}


As it has been described in the previous sections, the perturbations for the
$SU(2)$ field are governed by operators with a well-established resurgence
character such as the Lam\'{e} operator. The singularities analysis (using,
for instance, the Borel-Pade approximation) of these operators in the Borel
plane of the effective coupling constant is crucial in order to understand the
resurgent character of these type of perturbations. This will be done in this section.

On the other hand, although resurgence techniques in one-dimensional quantum mechanical systems
are very well tested, in the case of two-dimensional (and, indeed, higher
dimensional) potentials the situation is far less clear. Since the
electromagnetic perturbations satisfy an effective two-dimensional Schr\"odinger
equation, the spectrum of the electromagnetic perturbations must be
determined numerically. In this section we will obtain
numerical results relating the electromagnetic perturbations with the following two-dimensional Schr\"{o}dinger equation
\begin{equation}
-\frac{1}{L_{r}^{2}}\frac{\partial^{2}}{\partial r^{2}}\Psi(r,\theta)-\frac
{1}{L_{\theta}^{2}}\frac{\partial^{2}}{\partial\theta^{2}}\Psi(r,\theta
)+\biggl(\frac{k_{3}^{2}}{L^{2}}-4K\sin^{2}(\alpha)\sin^{2}(q\theta)\biggl)\Psi(r,\theta
)=\omega^{2}\Psi(r,\theta)\ .\label{schroodinger_main}%
\end{equation}

\subsection{WKB Analysis for the $SU(2)$ perturbations}

\begin{figure}[ptb]
\centering
\includegraphics[scale=0.3]{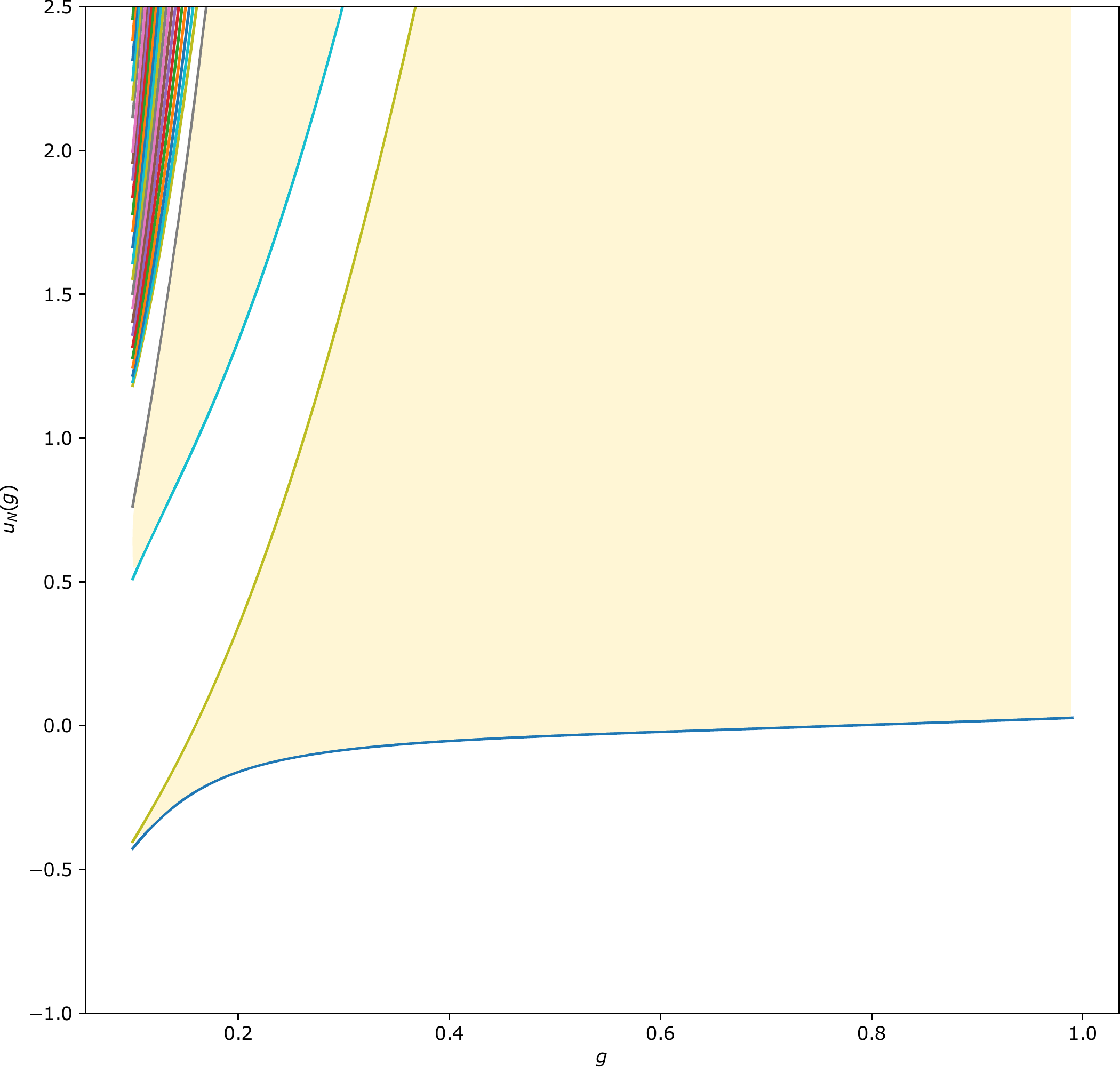} \qquad
\includegraphics[scale=0.3]{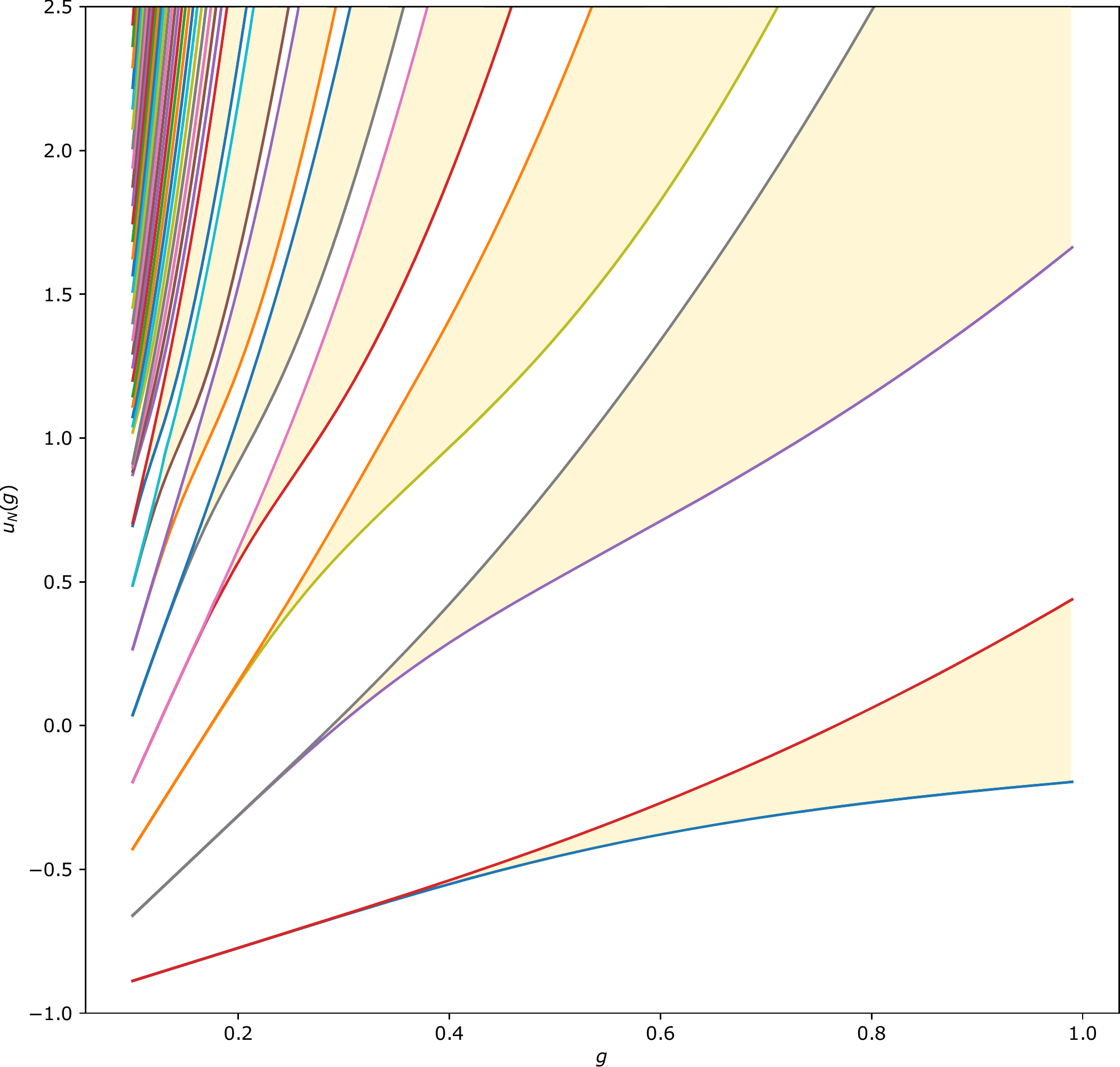}
\caption{Energy Spectrum for Lam\'e operators from $SU(2)$
perturbations for $m=0.2$ in (a) and $m=0.7$ in (b). The regions of stability
(the bands) are shaded and they are separated by regions of instability
(gaps), which are unshaded. We see a similar behaviour to the Mathieu spectrum
in wich at small $g$, the bands are exponentially narrow and high in the
spectrum, the gaps are exponentially narrow.}%
\label{fig3}%
\end{figure}

According to the previous section about the perturbations on the $U$ field, we need to study a Schr\"odinger equation of the form
\begin{equation}
-\frac{g^{2}}{2}\frac{d^{2}\psi}{dx^{2}}+V(x)\psi=u\psi \ ,
\end{equation}
where the potential is given by
\begin{equation}
V(x)=\operatorname{cn}^{2}\left(  x/k;k\right)  -\operatorname{sn}^{2}\left(
x/k;k\right) \ .
\end{equation}
We show the energy spectrum of our configurations as a function of the coupling
constant $g$ in Fig. \ref{fig3} (see Eq. \eqref{resurgent1}). For this purpose we employ the WKB method to obtain an
expansion in $g^{2}\rightarrow0$, namely
\begin{equation}
\Psi(x)=\exp\left(  \frac{i}{g^{2}} \int_{x_{0}}^{0} d x S(x)\right) \ .
\end{equation}
This is an standard procedure in which one solve the resulting Ricatti
equation using a power series ansatz for $S(x)$ as well as for the energy, that is%
\begin{equation}
S(x)=\sum_{n=0} g^{2 n} S_{n}(x) \ , \qquad E=\sum_{n \geq0} a_{n} g^{2 n} \ . 
\end{equation}
Here we have used the BenderWu package \cite{benderwu} to compute the WKB expansion obtaining a perturbative asymptotic expansion of the ground state energy (we
will not consider higher level states in this paper).
\begin{figure}[ptb]
\centering
\includegraphics[scale=0.5]{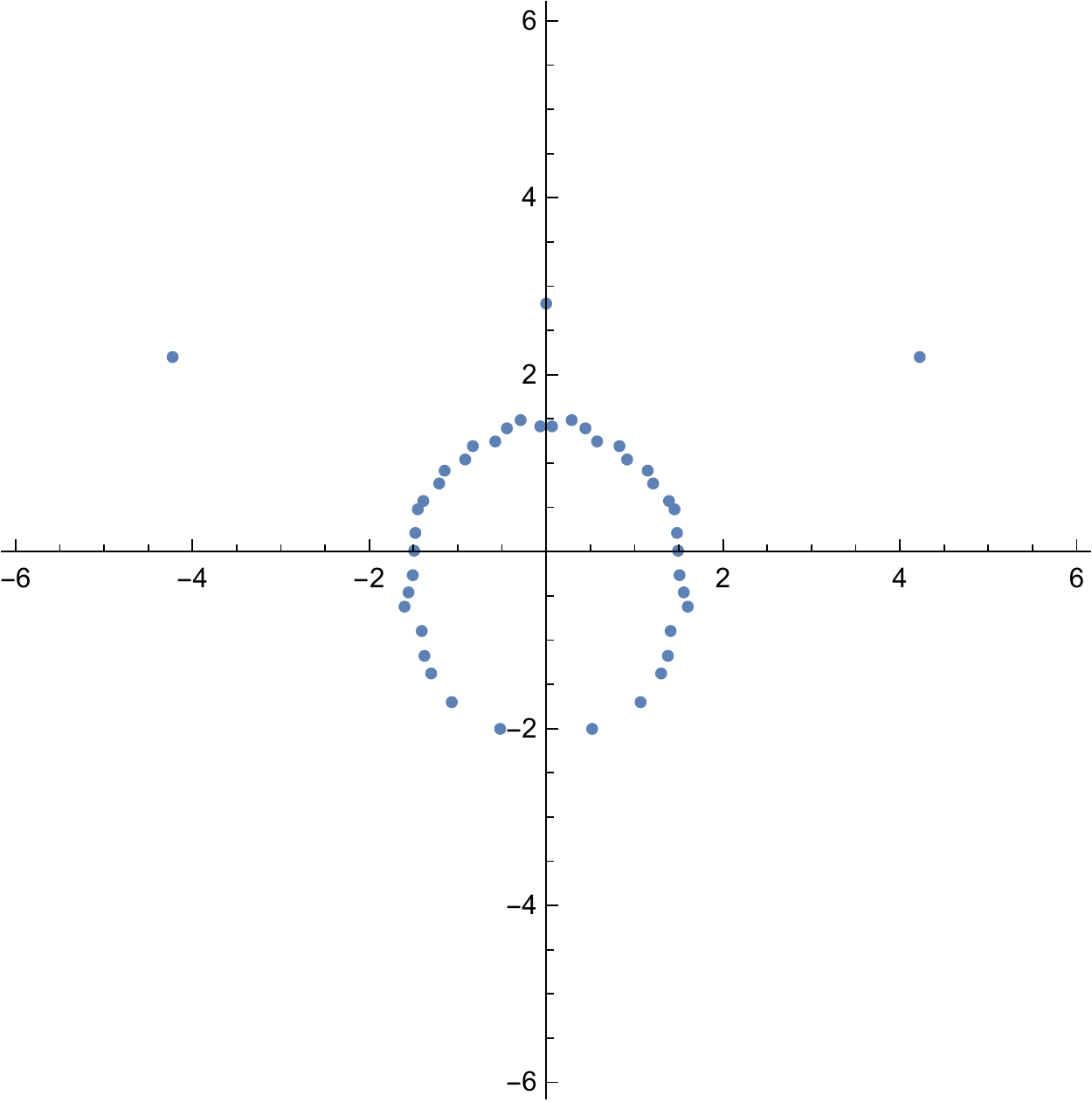}\caption{The complex
Borel plane for elliptic modulus $m=0.9$, dots indicating poles of the
Borel-Pad\'e approximation obtained from 100 orders of
perturbation theory in $g^{2}$ (hence we computed a total of 50 poles).
Accumulations of poles are anticipated to encode branch cuts in the full Borel
transform, and isolated poles are expected to be residuals of the numerical
approximation.}%
\label{borelplane}%
\end{figure}

\subsection{Numerical approach for the perturbed Maxwell equation}

\begin{figure}[ptb]
\centering
\includegraphics[scale=0.25]{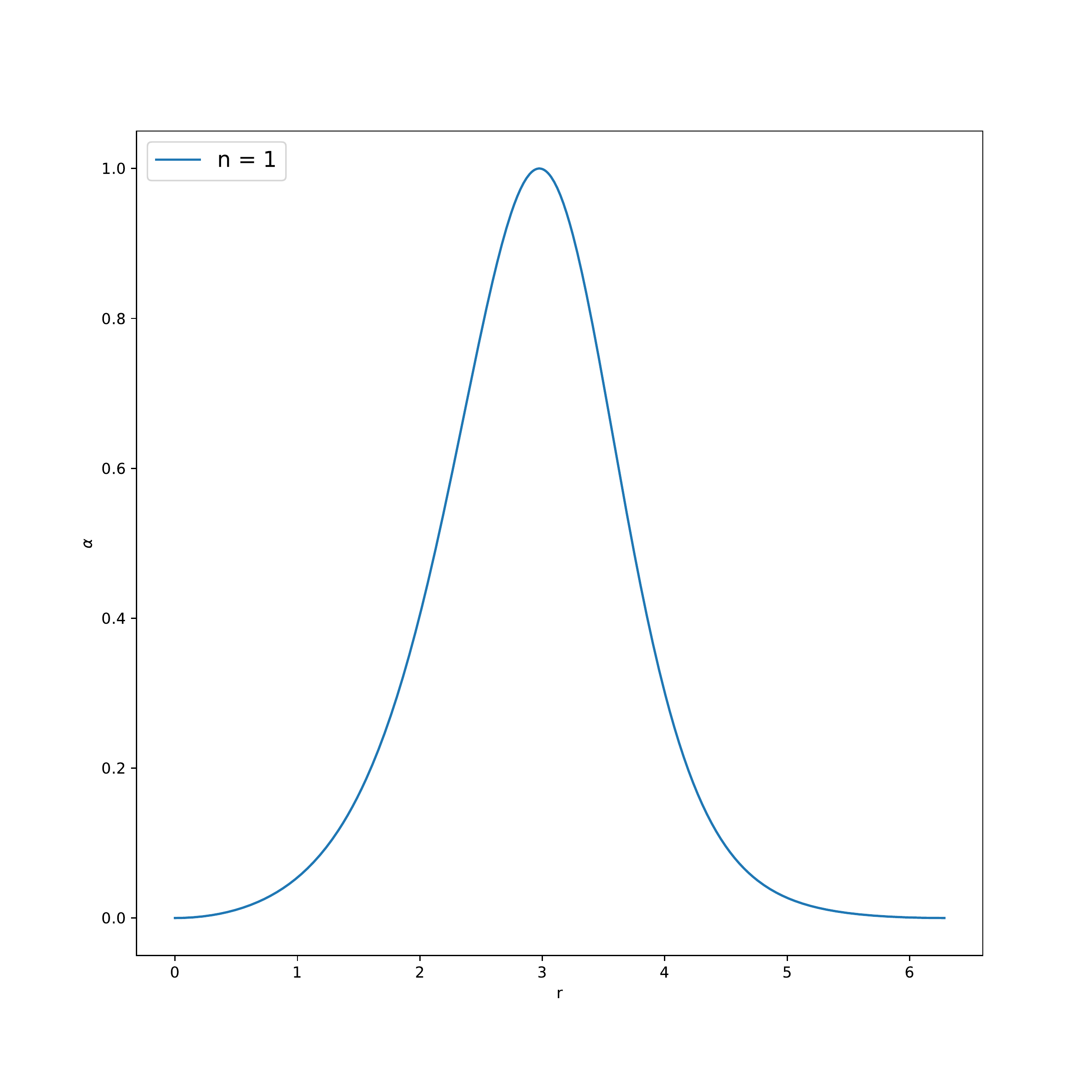} \qquad
\includegraphics[scale=0.25]{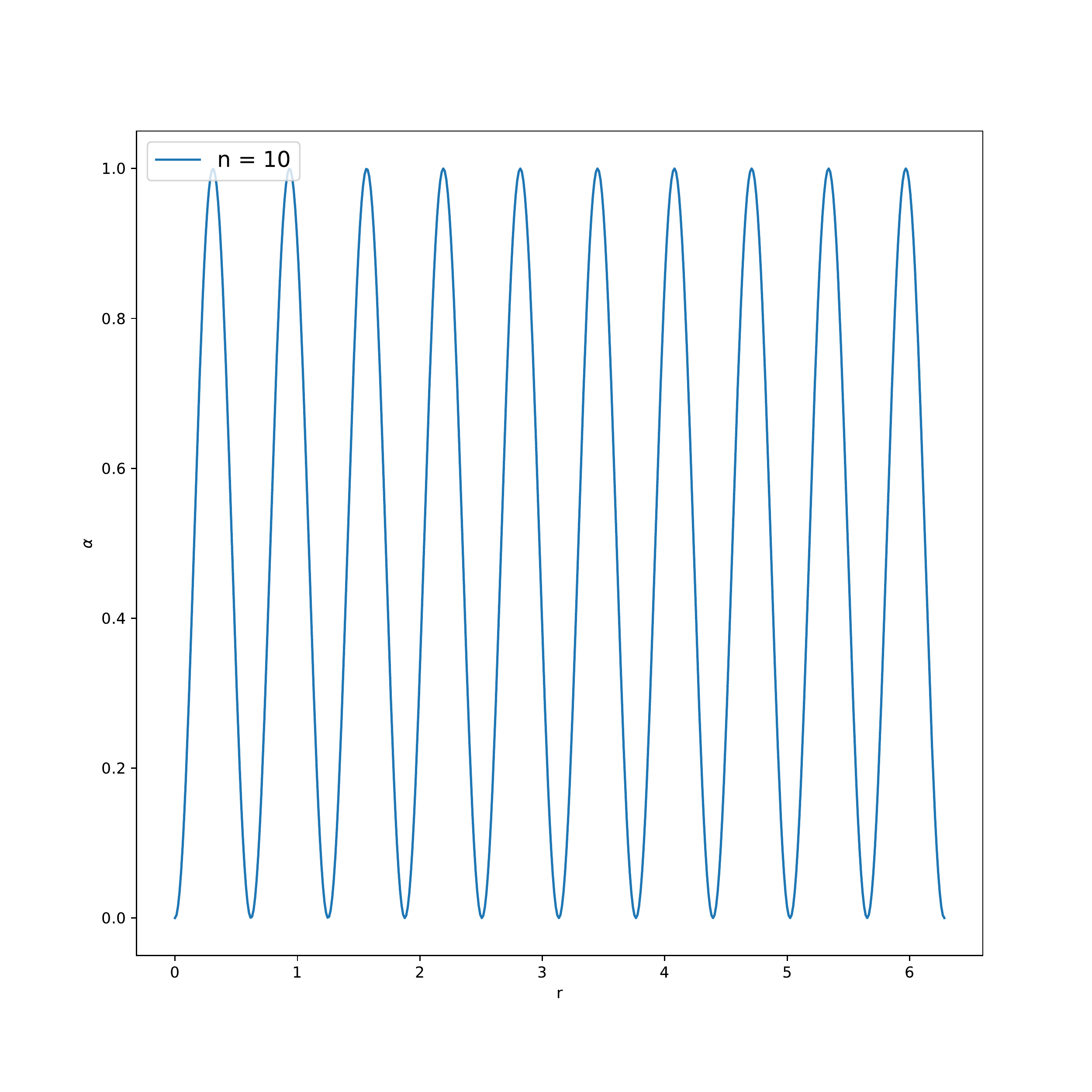}
\caption{$r$-component of two-dimensional potential for the perturbed Maxwell
equations: $\sin^{2}(\alpha(r))$ with $n=1$ and $n=10$.}%
\label{fig1:image2}%
\end{figure}

\begin{figure}[ptb]
\centering
\includegraphics[scale=0.3]{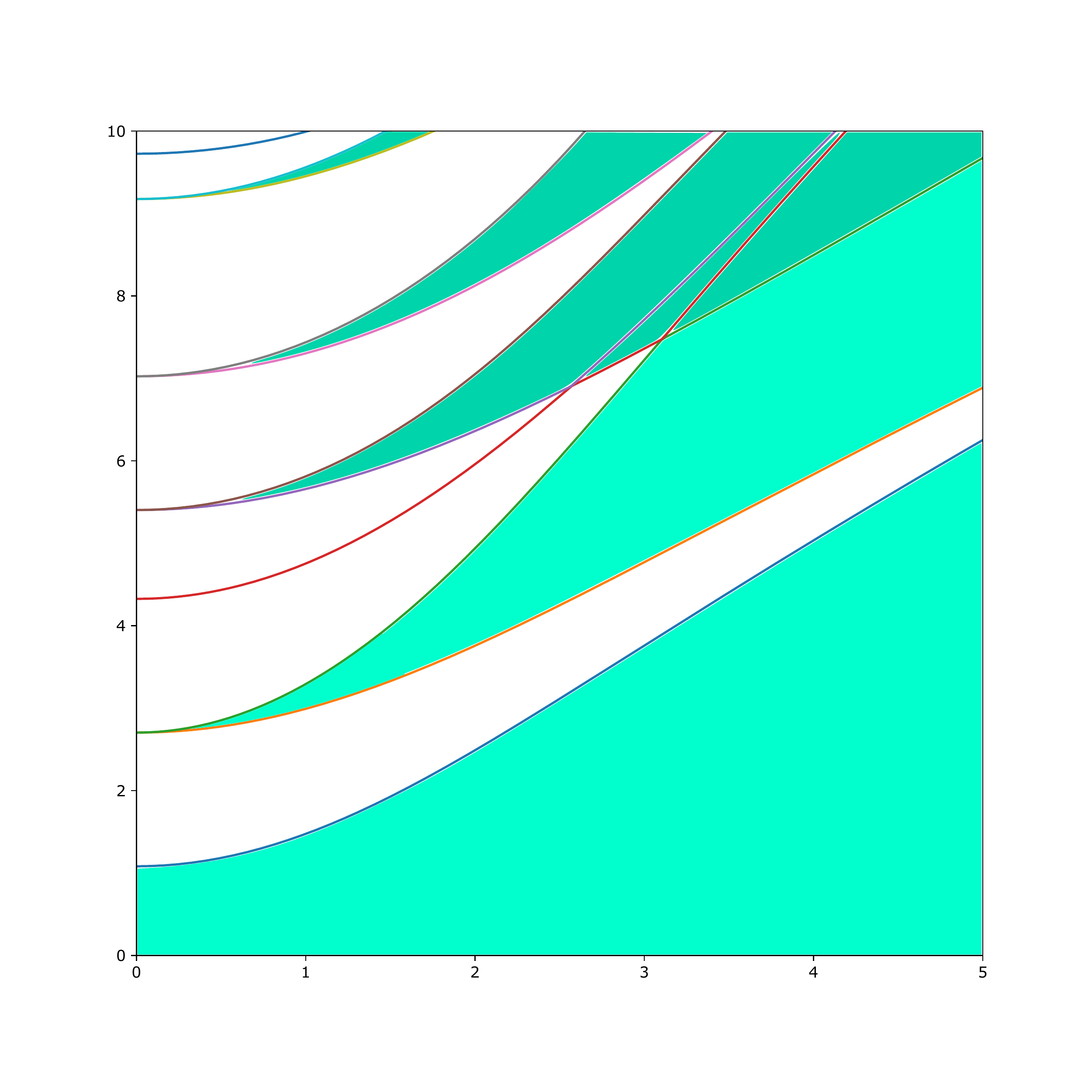}\caption{the Energy
Spectrum for the potential $L^{2}\sin{(\alpha(r))}^{2}\sin{(\theta)}^{2}$ as a
function of L, where $\alpha(r)$ is a numerical solution of Eq. \eqref{NLSM_n}
with $n=1$}%
\label{fig2:image2}%
\end{figure}

Here we present numerical solutions for the two-dimensional eigenvalue problem in
Eq. \eqref{schroodinger_main} using the Finite Difference Method, considering a
two-dimensional grid in the $(r,\theta)$-plane of $N_{r}\times N_{\theta}$
grid of $70\times70$ on a domain delimited by the ranges $0<r<2\pi$ and $0<\theta<\pi$.
We have used the Python Library \textit{Linear algebra} (scipy.linalg) for
eigenvalue problems. For the numerical implementation we will consider for simplicity
$L_{r}=L_{\theta}=L$. In this case the gauged NLSM field
equations and the perturbed Maxwell equations read
\begin{equation}
\alpha^{\prime\prime}-\frac{q^{2}}{2} \sin(2 \alpha)+\frac{4 m^{2}}{K}
\sin(\alpha)=0 \ , \label{NLSM_n}%
\end{equation}
\begin{equation}
-\left(  \frac{\partial^{2}}{\partial r^{2}}+\frac{\partial^{2}}%
{\partial\theta^{2}}\right)  \Psi+ \left(  \frac{k_{3}^{2}}{L^{2}}-4KL^{2}\sin
^{2}(\alpha(r)) \sin^{2}(q \theta)\right)  \Psi=E\Psi \ . 
\end{equation}
The boundary conditions for the profile $\alpha(r)$ according to Eqs. \eqref{bc}, \eqref{Bfinal} are $\alpha$ are $\alpha(2 \pi)-\alpha(0)=n \pi$, so that
the parameter $n$ in the boundary conditions is equal to number of ``peaks" of the potential (or energy density) in the $r$-direction of the lattice, as it
is shown in Figure \ref{fig1:image2}). The potential is fully determined by fixing the parameters $q$,
$K$ and $m$. In Figure \ref{fig2:image2} we plot the energy eigenvalues as
function of $L$. Making the comparison with the one dimensional Mathieu system
\cite{[17]} (where the band-gaps are modulated by $\hbar$) one can note that the parameter $L$ plays the
role of an effective $\hbar$.


\section{A comparison with the (1+1)-dimensional crystals in the Gross-Neveu
model}


The Gross-Neveu model (GN in what follows) is one of the most studied Quantum
Field Theory in (1+1)-dimensions since it shares many non-trivial properties
with interacting quantum field theories in (3+1)-dimensions, such as the
appearance of non-trivial condensates at non-perturbative level, asymptotic
freedom, Chiral symmetry breaking, dimensional transmutation and so on, but
can be analyzed with the tools of the integrable models. The GN model is a
theory of $N$ Dirac Fermions interacting via a quartic Fermionic potential describing by the Lagrangian 
\begin{equation}
\mathcal{L}_{\text{GN}}=\sum_{a=1}^{N}\left(  i\bar{\psi_{a}}\gamma^{\mu
}\partial_{\mu}\psi^{a}+\frac{g^{2}}{2}(\bar{\psi_{a}}\psi^{a})^{2}\right)
\ .
\end{equation}
This theory exhibits a phase diagram with a crystalline structure similar to
what is expected in QCD, which is invisible to standard perturbation theory
being only accessible to the non-perturbative $1/N$ expansion \cite{dunne1},
\cite{dunne2}, \cite{dunne3}. Using an auxiliary field, the GN field equations
can be written as a Hartree-Fock-Dirac equation (see, for instance,
\cite{thies0} and \cite{thies1}), that is
\begin{equation}
\left(  \gamma^{5}\frac{1}{i}\frac{\partial}{\partial x}+\gamma^{0}%
S(x)\right)  \psi(x)=\omega\psi(x)\ .
\end{equation}
The self-consistent analysis of the previous references allows to assume that
$S(x)$ is a Lam\'{e} potential leading to an analytical solution for the GN
model describing a crystal of kinks at finite density. The upper component of
the Dirac spinor $\phi_{+}$ corresponds to an Elliptic Function and satisfies
a Schr\"{o}dinger equation of the form
\begin{equation}
\left(  -\frac{\partial^{2}}{\partial\xi^{2}}+2\kappa^{2}\operatorname{sn}%
^{2}\left(  \xi|\kappa^{2}\right)  \right)  \phi_{+}=\mathcal{E}\phi_{+}\ ,
\end{equation}
where
\begin{equation}
\mathcal{E}=\frac{a^{2}}{\ell^{2}}\omega^{2}+\kappa^{2}\ .
\end{equation}
The eigenvalue term $\mathcal{E}$ depends on the constants $a$ and $\ell$, which
are fixed by the mean density. The elliptic parameter $\kappa$ is fixed by
minimizing the ground state energy density
\begin{equation}
E_{\text{gs}}=-2\frac{\ell^{2}}{a^{2}}\int_{k_{\min}}^{k_{\max}}%
\frac{\mathrm{d}k}{2\pi}\sqrt{\mathcal{E}-\kappa^{2}}+\frac{\ell}{2Ng^{2}%
a^{2}}\int_{0}^{\ell}\mathrm{d}\xi\tilde{S}^{2}(\xi)\equiv E_{1}+E_{2}\ .
\end{equation}
It is relevant to note the close similarity of the results in the references
\cite{thies1}, \cite{thies2}, \cite{thies3}, \cite{thies4}, \cite{dunne1},
\cite{dunne2}, \cite{dunne3} and \cite{SGkink} in the case of the solitons
crystals in (1+1)-dimensions with the (3+1)-dimensional gauged solitons
discussed here. Not only the energy density of the solitons crystal of the GN
model and of the Sine-Gordon model in \cite{SGkink} are very similar to the
energy density of the present gauged solitons. This similarity is especially
clear comparing the form of the reduced energy density defined in Eqs.
(\ref{ED1}) and (\ref{ED2}) with, for instance, the corresponding expression
in \cite{SGkink}. Also, the spectrum of the fluctuations of these
(1+1)-dimensional solitons crystals in \cite{thies1}, \cite{thies2},
\cite{thies3}, \cite{thies4}, \cite{dunne1}, \cite{dunne2}, \cite{dunne3} and
\cite{SGkink} are determined by a one-dimensional Schr\"{o}dinger operator of
the same Lam\'{e} family as the one in Eq. \eqref{lame2}. This supports very
strongly the existence of non-homogeneous condensates in the low energy limit
of QCD in (3+1)-dimensions.


\section{Conclusions and perspectives}


In this paper we have shown analytically that topologically non-trivial gauged
solitons of the (3+1)-dimensional gauged non-linear sigma model at finite
Baryon density are natural sources of Force Free Plasma. For these
multi-solitons most of the total energy and the topological charge are
gathered within tube-shaped regions while the magnetic field lines go around
the tubes. Our explicit analytical solutions allow to discuss the trajectories
of charged test particles moving close to these gauged solitons and also
identify the proper resurgent parameters through the analysis of the
perturbations. In particular, the perturbations of the solitons profile are
related to the Lam\'{e} operator with a suitable resurgent parameter. On the
other hand, the electromagnetic perturbations of the above system satisfy a
two-dimensional effective Schr\"{o}dinger equation, where the soliton's
background interacts with the electromagnetic perturbations through an
effective periodic potential in two spatial dimensions. Also we have studied
numerically the band energy spectrum for different values of the free
parameters of the theory and we found that bands-gaps are modulated by the
potential strength. Finally, we have shown that the crystal solutions
constructed here are qualitatively very similar to those of the Gross-Neveu
model in (1+1) dimensions, which strongly support the existence of
non-homogeneous condensates in the low energy limit of QCD in (3+1)-dimensions.

\subsection*{Acknowledgements}

F. C. has been funded by Fondecyt Grants 1200022. G. B. has been partially funded by FONDECYT Grant 1181047. A. V. is funded by FONDECYT
post-doctoral Grant 3200884. The Centro de Estudios Cient\'{\i}ficos (CECs) is
funded by the Chilean Government through the Centers of Excellence Base
Financing Program of ANID.

\end{document}